# IDN Authoring – a design case

Frank Nack

Any form of design[1] is a creative process considering goals and constraints that result in a product, which addresses aesthetic, functional, economic, or socio-political considerations that are established by the stakeholders involved. A design can follow two main development paradigms, namely a rational or action-centric model [1].
The rational model is plan-driven, where the plan is executed in terms of a discrete sequence of stages, such as pre-production (i.e. design brief, analysis of design goals, requirement specification, conceptualising solutions), production (i.e. development and testing), and postproduction (i.e. introducing the solution into the environment, evaluation and conclusion). This procedural view[2] allows for the repetition of stages at any time before, during, or after production based on accompanying best practices [2-7]. The rational paradigm understands design as informed by research and knowledge in a predictable and controlled manner.
The action-centric paradigm apprehends design also as informed by research and knowledge but considers the design process as improvised, where the sequences of analysis, design and implementation are undistinguishably connected [8-13]. Within action-centric design designers conceptualise a problem (framing) by defining goals and objectives and then construct tentatively the object to be designed (move). The design process here asks the designer to simultaneously refine the mental image of the design object based on the actual perception of the context in and for which it needs to be designed (sensemaking–coevolution–implementation framework [11]). The expression of the idea through suitable design tools facilitates the critical rethinking of the perceived idea, which results in a new design cycle. The essential element in action-centric design, the development of any design idea, is the means of expression and perception to facilitate a space for inspiration, ideation, and implementation through which the designer refines ideas and explores new directions [12]. This way of looking at design is known as design thinking [13].
Both paradigms are essentially human-centred, in the way that they see human needs, competences (cognitive as well as knowledge) and behaviour as the basis of the investigation and only then design to accommodate those [4]. At its core, all design is thus experiential and that is why all design addresses fundamental concepts, such as affordance, mapping, feedback, conceptual models, and system image. Affordance refers to the relationship between the properties of an item and the capabilities of an agent that establish how that item can be used or understood. Mapping represents the relationship between the elements of two sets of items (i.e. the layout of controls and the devices being controlled). Mappings represent analogies, which can be spatial, biological, cultural, or follow principles of perception (such as Gestalt psychology) to facilitate grouping through proximity. Feedback communicates the result of an action and in design it is usually linked to the human nervous system, i.e. the visual, auditory, touch and olfactory sensory systems, and, depending on the design domain, also with the proprioceptive and vestibular systems. Within human-centred design the feedback is immediate and informative. Conceptual models, in the context of humans, represent mental models that can address the individual itself, other humans, the environment, and objects, or services the individual or a group of individuals interact with. The models can be formed by experience, through training or instruction but they are necessary to facilitate people to understand the world. With respect to design this is relevant because everything that is designed needs to communicate through its physical structure and appearance what it is good for and how it can be used. Norman calls this distinctly object/system relation the system image, which is reflection-based. “The user’s conceptual model comes from the system image, through interaction with the product, reading, searching for online information, and from whatever manuals are provided. The designer expects the user’s model to be identical to the design model, but because designers cannot communicate directly with the users, the entire burden of communication is on the system image.” [4, p.31].
In whatever way one looks at it, design can be brief or lengthy and complicated, involving considerable research, negotiation, reflection, modelling, interactive adjustment and re-design, and it is based on good conceptual models, which in turn require good communication.
In this text, we consider the authoring of an interactive digital narrative (IDN) as a system of interwoven creative processes and hence can look at it as a design process. The aim is to better understand the structural, aesthetical and interactive concepts authoring must address, how authors think about those and what that means regarding the tools required to support authoring for IDNs. We first look, therefore, at what needs to be designed, then how it should be done and then what type of tools are required.

---

[1] The closest design fields we relate to in this chapter are communication, interaction, experience, and process design. Industrial design is potentially relevant if an interactive digital narrative (IDN) is considered as physical space.
[2] Which is manifested in process models such as the waterfall model, or the systems development life cycle.

Narratology

Narratology is the study of narrative and its structure and the ways how narratives effect human perception, cognition, and emotion. Each story told, whatever culture, is the reconstruction of experiences in narrative form, and these become the units of remembered life. A narrative, finely tuned or rough, can be presented in all forms of human expression, as in speech, literature, image, film, music, song, theatre, dance, pantomime, or game.
The earliest form of western-oriented narratology[3] can be found in Aristotle's Poetics [15], which provides in mimesis[4] the all-encompassing concept based on which six constituent elements of narrative arise: plot, character, language, thought, spectacle and melody [16, vol1, p. 33 -34]. Modern narratology starts with the Russian Formalist movement (1918 - 30), that developed the notion of an opus as the sum of all applied artistic skills and in this context focused on understanding and describing these skills in their respective functionality (practical, theoretical, symbolic and aesthetic) within prose, metre, and literature evolution [17, 18 19[5]]. The philosophical context of Russian Formalism exerted great influence on the development on structuralism. Structuralism formulated a rationalised and deductive approach to narration, which considered narrative structure as analogous to language structure and thus linked structure with the determination of content [20 – 26]. An analogous approach to the representation of narrative structure, in the field of Artificial Intelligence, considered the applicability of story grammars to text understanding [ 27 - 32], where the main influences came from Propp's work on Russian folktales [18] and Chomsky's transformational grammar [20]. Related concepts like discourse [33], conceptual dependencies [34], scripts, plans, goals [35], dynamic memory [36] and schemata [37] appeared around the same time. The latter three are essential because they address crucial elements of narration, i.e. memory that facilitates the accumulation of knowledge about how the social world around us works, and expectation about how events should unfold. Both are necessary processes that enable easier comprehension of a complex world, exactly one reason why narratives are told. Post-Structuralism in form of reader-response criticism[6] disagreed with the focus on structure and argued in favour of the reader's interpretive activities. This school of literature theory argued that the reader is an active agent who conveys through personal interpretation the real in a work. Here interpretation is considered as a performing art in which each reader creates the own, possibly unique, text-related performance and understanding. In this view reading is always both subjective and objective. This point of view also alters the role between author and perceiver, as Barthes stated: '...a text is made of multiple writings, drawn from many cultures and entering into mutual relations of dialogue, parody, contestation, but there is one place where this multiplicity is focused and that place is the reader, not, as was hitherto said, the author. The reader is the space on which all the quotations that make writing are inscribed without any of them being lost; a text's unity lies not in its origin but in its destination.' [38, p. 148]. In this context, the concept of anticipation is relevant, which includes expectations, frustration expectations, retrospection, and reconceptualization of new expectations. Iser outlines this as: "We look forward, we look back, we decide, we change our decisions, we form expectations, we are shocked by their nonfulfillment, we question, we muse, we accept, we reject; this is the dynamic process of recreation." [39, p. 288]. What all these approaches unites is "…the recognition that narrative theory requires a distinction between "story," a sequence of actions or events conceived as independent of their manifestation in discourse, and "discourse," the discursive presentation or narration of events." [40, p 189].
The problem is the diverting introduction of terms and hence meaning in narrative terminology. The Russian Formalists suggested the doublet fabula and syuzhet. Other proposed pairings are histoire/discourse, histoire/récit, or story/plot for this dualistic phenomenon. The different terms are not equivalent in use, and it depends which narratology tradition one follows, namely the thematic (a semiotic formalization of the sequences of the actions told) or modal narratology (examines aesthetic aspects, such as stressing voice, point of view, the transformation of the chronological order, or rhythm) [41]. This text follows the argument by Sternberg [42] and Ricoeur [16] that thematic and modal narratology should not be looked at separately. As the aim is to shed light on the authoring process for IDNs from a design point of view, it seems also not fruitful to use the above doublets due to their semantic fragility. Thus, here the doublet narrative and narration will be applied.

[3] There are naturally other narrative models and traditions. Kishōtenkets, for example, describes the structure and development of classic Chinese, Korean and Japanese narratives. For the importance of this cross-cultural view for the design and understanding of narratives in general and for IDNs in particular see [14].

[4] Aristotle considered mimesis as the "an imitation or representation of action" in the context of narrative representation. Mimesis must be understood in the context of a certain distance between the work of art and life on the other, as without this distance, tragedy could not give rise to catharsis. At the same time the work of art has to make the audience to identify with the characters and the events, as otherwise it does not touch the audience. Aristotle holds that it is through "simulated representation," mimesis, that we respond to the acting in the work of art. It is the task of the creator to produce the enactment to accomplish this empathy.

[5] With a reference Tomashevsky's "Thematics"

[6] Representatives include Norman Holland (psychological reader-response theory, individual reader-response theory), Stanley Fish (Affective stylistics, social reader-response theory, individual reader-response theory), Wolfgang Iser (Uniform reader-response theory), Hans-Robert Jauss (uniform reader-response theory), and Roland Barthes. The reader-response approach is also applied in cinema (David Bordwell), or visual art (E. H. Gombrich).

## Narrative and narration

So, what is it that needs to be designed? Let us start with the narrative part.

A narrative, in short, represents events that '... are arranged and connected according to the orderly sequence in which they are presented...' [19, p. 67]. It embodies any account of related events or experiences, organised as a representational system based on surface structures (expression) and deep structures (content), where each of those distinguish between substance and form. Substance, in this context, represents the natural material for content and expression, whereas form represents the abstract structure of relationships a particular media demand [43]. Figure 1 shows the relationships between the differing structures found in a narrative.

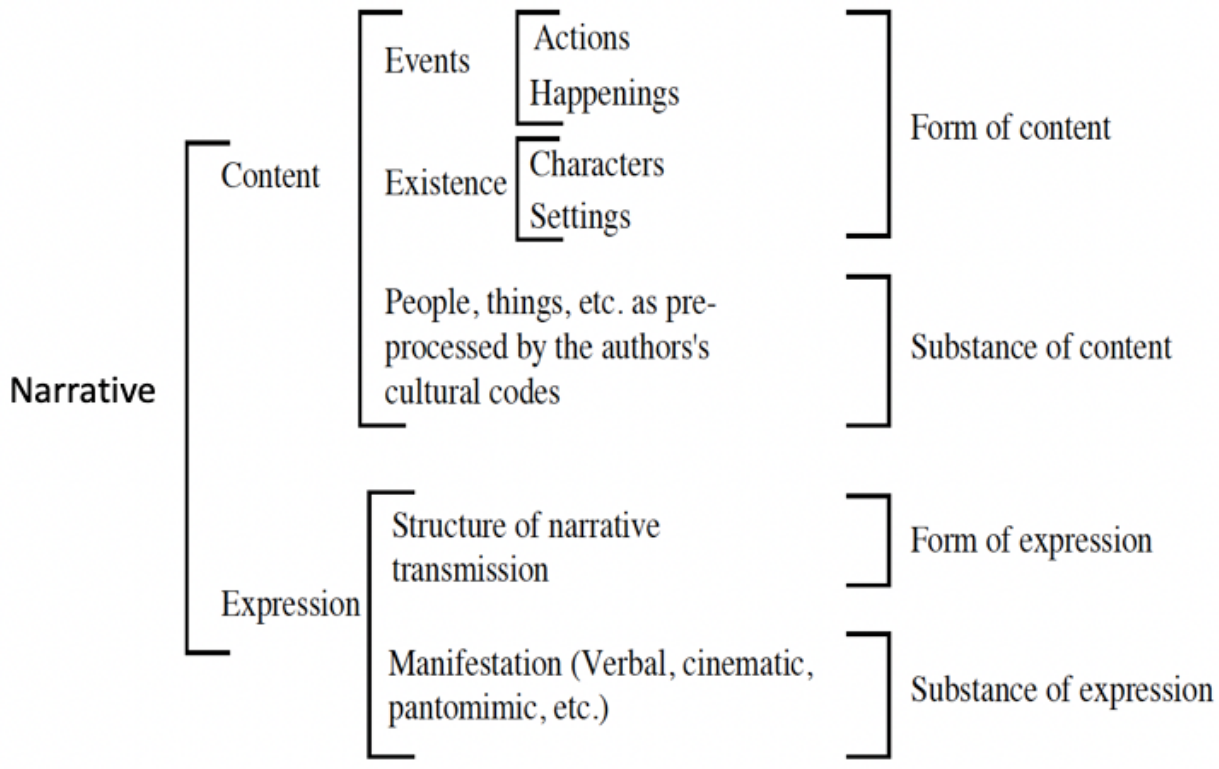

**Figure 1** Relationship between narrative elements (adopted from Chatman [43, p.26])

The actual narrative, understood from the point of view of the traditional linear narrative, is a final media artefact, i.e. the text, the film, the performance, that has been shaped by the creator[7], particularly about the plot (the series of events), to deliver information to the audience[8]. Here the creator makes use of narrative techniques, such as the setting (the time and geographic location within a narrative), the plot (e.g. backstory, analepsis or prolepsis, foreshadowing, in media res, twists, red herring, deus ex machina, love triangle, story within a story), perspective (e.g. audience or author surrogate, the fourth wall, multiperspectivity, first/second/third person narration), style (e.g. allegory, balthos, hyperbole, metonymy, metaphor, pathos, satyr), themes (e.g. irony, love, grief, lust, guilt), genre (e.g. romantic love story, science fiction, western, screwball comedy). This shaping is guided by intellectual mechanisms or skills, namely prototyping (organise the identification of types of persons, actions, localities, etc.), templating (articulate common story formats, where each formal element represents a structural story movement, realising the stages through which the agent of the story must pass, such as: orientation, complication, evaluation, resolution, coda), and procedural thinking (organise the search for appropriate motivations and relations of causality, time and space) [45 – 47].

At the same time, a narrative is a communicational system[9]. Narration, or storytelling, means making a comment about a certain event, following an idea about the medium and form of presentation which is grounded in one's own motivational and psychological attributes. In this line of thinking a narrative represents a dynamic psychological entity that refers to mental or conceptual objects such as themes, goals, beliefs, or assumptions. The dynamicity of the narrative is already manifested in the construction process, as narration is a targeted phenomenon. There is a receiver and the creator’s perception of him or her will doubtless have an impact on the outcome of the narrative to be told and the way how it is told. In fact, the dynamics within narrative construction are twofold. On the one hand, the intentions of the creator must be achieved, i.e. to present the content material

[7] We use in this context the term creator to not mix it up with the term “narrator”, as this represents the actual teller of the story (i.e. first-, second-, third-person, or alternating person point of view [16]). We also do not want to use the term “author” as this indicates the creator or originator of any written work.

[8] The two most influential narrative structures in western narrative tradition are Aristotle’s Poetics [15] and Campell’s Monomyth/The Hero’s story [44]. Yet, those are not the only possible structures as outlined in [14]. Also have a look at Wolfgang Walk’s view on why the Hero’s story structure does not necessarily make sense, in the context of gaming (https://www.gamedeveloper.com/design/the-myth-of-the-monomyth). It will be shown later why this is of relevance for authoring.

[9] Narration is one of the four discourse modes of language–based communication, besides argumentation, description, and exposition [48, 49]

as plausibly and succinctly as possible so that the aimed for intellectual as well as experiential and psychological impact can be achieved. The effect of this relies on articulation techniques, i.e. communication strategies between creator and receiver. A narrative can, for example, be organized into a number of formal categories: nonfiction (such as creative non-fiction, biography, journalism, transcript poetry and historiography); fictionalization of historical events (such as anecdote, myth, legend and historical fiction) and fiction proper (such as literature in the form of prose, short stories, novels, narrative poems and songs, and imaginary narratives as portrayed in other textual forms, games or live or recorded performances), of which each performs in its idiosyncratic communication constraints. On the other hand, the dynamics within the material must be considered, since its affordances form the bases so that the narrative can be delivered. Those material dynamics are described in principles and procedures for speech [15] [10], literature [50, 16], theatre [51 - 53], film [54, 55], comics [56], dance [57,58], music [59], or game [60].
Research in discourse psychology further supports the above statements. Discourse psychology established process models in form of coherent network of interrelated propositions that reflect the explicit information within a narrative, and the inferences that establish how the particular content is interrelated to different forms of knowledge, i.e. situated, partial, domain, and procedural (skill) knowledge [61 - 64]. The established models, for example, show that narrative receivers monitor situational dimensions, update the mental model when dimensional changes occur, and hence arrange events around these dimensions in their episodic memory. Other findings are that agents, objects, and abstract concepts are entities that prominently contribute to narrative structure [65], that narrative events are connected via causal temporal and spatial relationships that are hierarchically organised goal episodes [62, 66] and that the multidimensionality of mental models might be equally important across media [67]. It seems that time is more important in conveying shifts between narrative events than space, no matter what medium is used. Moreover, findings indicate that perceptual and cognitive processes that operate during narrative understanding can be diverted into frontend and back-end processes. Front-end processes address the moment-to-moment information processing, whereas back-end processes address the building and maintenance of mental models and experiences. As different media still require special affordances with respect to information distribution (i.e. use of conventions) as well as the receiver's comprehension literacy, it has been suggested that there are trans-symbolic (i.e. universal and operational across media) and symbolic-specific process (i.e. unique to a particular medium), where front-end processes affect the information that feeds back-end processes [68,69]. This empirical work also aligns with semiotic sign systems (i.e perception, textual, social, and syntagmatic and paradigmatic codes) and signification (denotation of signs and kines as its first level and additional 2 connotation levels – see [70]). This combination indicates why control over content is possible. Along those lines develops work in cognitive narratology [71,72], which addresses the conceptualisation of mental representations of time, space and cause-effect and the elements of narrative that enable authors and perceivers to construct these, such as schemas and frames. They address the mental relations between perceiver and the narrator and character, for example via point of view, dialogue, description of action and body language and conveyance of emotion. Empirical studies on these cognitive processes have been conducted by [73, 74]. In relation to authoring see for example [75].
Finally, both creator and receiver do not exist in a vacuum but share a social environment which adds extra structures to the narration process. Every culture has its own narratives, which are shared for different reasons, such as teaching moral values or to entertain. Groups of such tales merge into story cycles (see the Arabian Nights [76,77]), or clusters of fairy tales, folktales, mythology, legends, fables up to more modern representational forms of history, personal, political and documentary narratives. A culture's "stories create a shared history, linking people in time and event as actors, tellers, and audience." [45, p.28]. A narrative is about meaning and hence, every new story adds to the stream of continually reconstructing the past to influence, in the best way, the future. In that way, narrative and narration work in the same realm as design thinking, which is called in narrative terms "narrative mode". Narrative mode is concerned with human wants, needs, goals, and intentions, and facilitates the audience to observe characters in their actions and to realize which obstacles were encountered and which intentions were realized or unfulfilled [78].
This aspect of observation and realisation over time, or as Jakobson described it as the application of combination and selection in the operation of a sign system resulting in a system of meaning based on alternations and alignments [79], makes the digital narrative a special design case[11], as it highlights the relevance of expectation as a crucial element of experience. Traditional linear narrative products, though tremendously successful and influential, can address narration in a restricted way because the creator can only estimate the expectations on the whole product as well as expectation levels during the perception process the audience undergoes while linearly accessing the narrative[12]. Adapting the design point of view by Preece et. al [6], the creator of a digital narrative mainly designs for an experience and not the experience itself. This problem is addressed in the development of interactive digital narratives (IDN), where the narration is also understood as a dynamic and complex process of interaction in a partly given social context, where 'the interaction encompasses on the communicator, the content, the audience and the situation' [80, p. 209] but where the emphasis of perception, expectation handling and hence meaning making is shifted to the audience side, very much adapting post-structural thinking.

[10] The Aristotelian model is mainly one option. It is mentioned here because it is well described. Other oral models and traditions for different cultures also follow particular canons and presentation norms, for example Bard (Celtic), Ashi (Turkic), Pingshu (Chinese), Dastangoi (India), Kobzar (Ukraine), Seanchai (Gaelic),Griot (West Africa), or Maggid/Minstrel (Hebrew/Jewish).

[11] All types of design address issues of selection and expectation but the design of a narrative or narrative space is broader with respect to the system image. That is why in this text the related design fields considered are rather communication, service and interaction design, than product or industrial design.

## Interactive Digital Narrative

A digital interactive narrative is a form of interactive media[13], only that the actual narrative is not the final product or service. IDNs lean towards the idea of a narrative environment in form of a cybernetic system[14] [83], that establishes a circular causality, where the observed outcomes of actions are taken as inputs for further action in ways that support or disrupt a condition. Condition in the context of a narrative should be understood as the development of narrative states over time.
Janet Murray established three key concepts that an IDN, which she understands as a participatory medium, must support, namely immersion, agency and transformation [87, pp 97 – 182]. Immersion represents the experience of the sensation of being enclosed in a completely other reality that takes over all the attention by occupying the whole perceptual apparatus. For that it is necessary to not only be willing to suspend disbelief but to actively create belief[15]. Where immersion focuses on the environment, agency provides the means to take meaningful action and see the feedback in form of effects or consequences resulting from the choices made. Murray merges here two properties of an IDN - the procedural and the participatory. Agency covers the engagement of the participant to a world that responds to this engagement in an expressive and coherent way [88]. The practice of narrative agency is the challenging point of authorship. In her opinion, an author should be seen as a choreographer who supplies the context, the feel for what will be performed. The interactor, on the other hand, takes the role of a protagonist, explorer, or builder, who makes use of the provided media and interaction means to improvise a particular performance within the digital story system. Thus, agency is the concept that requires a distinction between the derivate authorship of the interactor against the originating authorship of the system. Transformation focusses on the shape-shifting abilities of computational environments. Transformation here reflects the fragmentation powers of the machine, which facilitate the multidimensional presentation, rearrangement, spatial and temporal organisation of media mosaics, that finally support the power of juxtaposition for reflection[16]. A second way for her to look at transformation is the ability of some technology, in particular virtual reality (VR), to enact or construct owned stories of a set of formulaic elements (such as Propp's abstract narrative structures and archetypes). Here transformation relates to the impact of enactment within an immersive environment in form of a cyberdrama, which can establish a catharsis[17] effect, useful in entertainment as well as in psychotherapy. In the context of cyberdrama she also discusses the issues of open-ended narratives (refused closure [87, pp. 173 - 175], which works against traditional narrative expectations and can also result in moments of fear (disturbing content or break of imersiveness by returning into reality). In this context of never-ending, ever-morphing narrative spaces she also speculates on the role of tragedy.

---

[12] This does not mean that a linear product (i.e. a book or a film) cannot be presented in a non-linear fashion. Examples are: James Joyce's Ulysses (1922), Jorge Luis Borges' The Garden of Forking Paths (1941), or Italo Calvino's The Castle of Crossed Destinies (1973). Example films are Resnais's 1993 Smoking/No Smoking (1993), Tree Colours trilogy by Kieślowski (1993 – 94), Tykwer's Run Lola Run(1998), Shear's Urbania (2000), C. Nolan's Memento (2000), Daldty's The hours (2002), or Iñárritu's Babel (2006). Despite the structural and point of view complexities all exampled at the end provide one product in which the audience has to follow the choices made by the creator.
[13] In this context "new media" should also be mentioned, which covers media that are computational and rely on computers and the Internet for redistribution, such as games, websites, virtual worlds, or human-computer interfaces [81, pp. 13–25; 82].
[14] This also includes the extended view of second-order cybernetics [84, 85] and radical constructivism [86, pp. 17-40], which both look at information and knowledge from the point of view of observing systems (second order) rather than observed systems. The link between second-order cybernetics and radical constructivism is established though the view of knowledge as "situated knowledge" as a correspondence between a knower's understanding of their experience and the world beyond that experience.
[15] This covers action control (i.e. by providing detail), deign of the digital self, structured individual as collaborative participation, arousal regulation.
[16] She names this kaleidoscopic narrative, adapting McLuhan's observation, that communication media are mosaic rather than linear in structure [87, pp. 155 – 162].
[17] Catharsis is a term in dramatic art, introduced by Aristotle's Poetics [15], that describes the purifying and purging effect of tragedy and comedy (and quite possibly other artistic forms [89]) on the audience but as Aristotle is vague it can be considered that the approach also works on characters in the drama as well.

The outlined concepts can be found in various incarnations of IDNs. An intuitive and mono-media approach towards interactive narrative is ergodic literature [90], a linear text and a machine capable of producing several manifestations of a text. This type of interactive narratives is heavily influenced by a traditional view on fiction in a written form. The simplest form of this type of narrative is hypertext[18] [91 – 93], which represents networked nodes and depending on the choices the reader makes, the story evolves. There are often several options in each node that directs where the reader can go next, where the nodes can be organised in different ways, i.e. axial, arborescent, and networked. Those organisational forms are not exclusive within a piece of hyperfiction. Hypertext, though interactive, does not leave space for expectations and experiences on an individual perception level as the explorative space is designed by the author.

Cybertext can be considered as an extended form of hypertext as it tried to leave the author provided narrative structure of hypertexts in favour of the provision of software code to control the reception process without reducing interactivity [94].

Considering an interactive narrative as a communication system that contextualises relations between the self and other, private and public, inner thought and outer world, various application forms have been established to achieve the relevant form of information propagation [95, 96]. There is a rich collection of complex media narratives[19] already starting in the early 1990[20], which cover different types of narrative genres, techniques, styles, modes and presentation types to facilitate the exploration of complex problem spaces[21].

More complicated have been the developments in the direction of the cyberdrama. The essential theoretical work for this direction has been the book "Computers as Theatre" by Brenda Laurel [104], which was based on her Ph.D. dissertation from 1986. She was one of the first researchers who pointed out that functionality is not all computers and interactive systems can provide. She formulated an aesthetic, dramaturgic take on interface design by arguing that dramatic theory (Aristotle) is well suited, as it provides a mimetic approach to enable the process of enactment in form of action rather than description. In her book, she especially emphasises a first-person view as a condition for engagement in interactive systems and demonstrates this with examples on games, such as the Monkey Island adventure game series[22]. Around this time research had been performed on necessary technical infrastructure, first visually simple versions [105 – 110]. Further graphically advanced examples that also addressed more narrative dramatic structures [111] or serious problems in the context of role-play, such as harassment [112, 113], have been presented later. A different approach is offered by the *Façade* project [114], which uses natural language processing and other artificial intelligence routines to direct the action and interaction between the characters and the player. The overall architecture of such systems has three parts: a drama manager, an agent model, and a user model. The drama manager supervises the narrative flow by searching and executing story "beats" in a coherent sequence and refines story events by providing new information and resolving contradictory plot lines. The agent model collects information about the story world and characters and generates possible actions in response for each non-player character in the story. The user model keeps track of player choices and inputs, so that the drama manager and agent model can cooperate with the way the audience attempts to interact. The most popular incarnations of those type of story environments are massively multiplayer online games (MMOG or MMO) or the massively multiplayer online role-playing games (MMORPG)[23]. Current new developments on hard- and software supposed to lead to the Metaverse[24], which, due to its focus on social connection and interaction, might point to new forms of digital interactive narratives.

---

[18] Example hypertexts are: Michael Joyce (1990) afternoon, a story, Stuart Moulthorp's Victory Garden (1992), Douglas Cooper's Delirium (1994), Shelly Jackson's Patchwork Girl (1995), Adrienne Eisen's Six Sex Scenes (1995), Stuart Moulthrop's Hegirascope (1997), Caitlin Fisher's These Waves of Girls (2001), or Stephen Marche's "Lucy Hardin's Missing Period" (2010). See also The Electronic Literature Organization (ELO) that facilitates the writing, publishing, and reading of electronic literature (https://eliterature.org/). Myst (1993), a graphic adventure puzzle video game, could also be considered a hypermedia product, as it was built with the Hypercard software.

[19] The debate surrounding narratology and ludology will not be discussed in this chapter (see [97 – 100]. The view shared in this text is that games should be looked at for their stories and also be considered as interactive narratives (some at least) because they are participatory. See for example [101 – 103]

[20] See for example the works created at the Interactive Cinema group at the MIT Media Lab https://ic.media.mit.edu/

[21] More recent examples are Fort McMoney (2013) https://docubase.mit.edu/project/fort-mcmoney/, Last Hijack Interactive (2014) https://docubase.mit.edu/project/last-hijack-interactive/ , Voyage (2015) https://venturous.se/, Pregoneros or Medellin (2015) https://docubase.mit.edu/project/pregoneros-de-medellin/ , A dictionary of the revolution (2017) https://amirahanafi.com/post/155352117400/a-dictionary-of-the-revolution-2017 , Gaming for Peace: Kusam (2019) https://gap-project.eu/, The Industry (2020) https://theindustryinteractive.com/home

[22] https://en.wikipedia.org/wiki/Monkey_Island

[23] Examples are: Castle Infinity (1996), EverQuest (1999), The Martrix Online (2005), the most successful and best-know World of Warcraft with its eight expansion packs (2004), or Fortnite (2017). In this context, one could also consider sandbox games such as Minecraft (2011) or more feasible the Sims (2000), a life simulation sandbox game that features open-ended simulation of the daily activities of one or more virtual persons.

[24] The term "metaverse" originated in Stephenson's novel Snow Crash (1992) and covers the Internet as a single, universal immersive virtual world (including multisensory extended reality, and simulated reality) that can be accessed through the use

Research on narrative generation, discourse psychology on story comprehension and representation, as well as work on cognitive narratology establish a basis of models to describe what an IDN is. The most current definition is provided by Koenitz [116, p 16]: “Interactive digital narrative is a narrative expression in various forms, implemented as a multimodal computational system with optional analog elements and experienced through a participatory process in which interactors have a non-trivial influence on progress, perspective, content, and/or outcome.”.
In this definition, a systemic view on IDN is hidden, which Koenitz outlines in his SPP model (System, Process, Product) [117, 118], the currently best model to capture the fragmented, distributed and interactive nature of an IDN, which at the end results in a coherent presentation. SPP is informed by cybernetics and cybernetic art theory. In this model a ‘story’ – as understood by literary theory – is not manifest in the digital work, the system, which contains the protostory, a term describing all the potential narratives. Those can be instantiated by an interactor through an interactive process, which leads to a realized, reordered product (the actual story). Koenitz, following cognitive narratolog, understands a narrative as a mental frame which can be evoked by many different manifestations and discourse strategies. For the interactor, the process is a double hermeneutic circle in which they reflect both the instantiated narrative path and the possibilities for future interaction [118], as outlined in Figure 2.

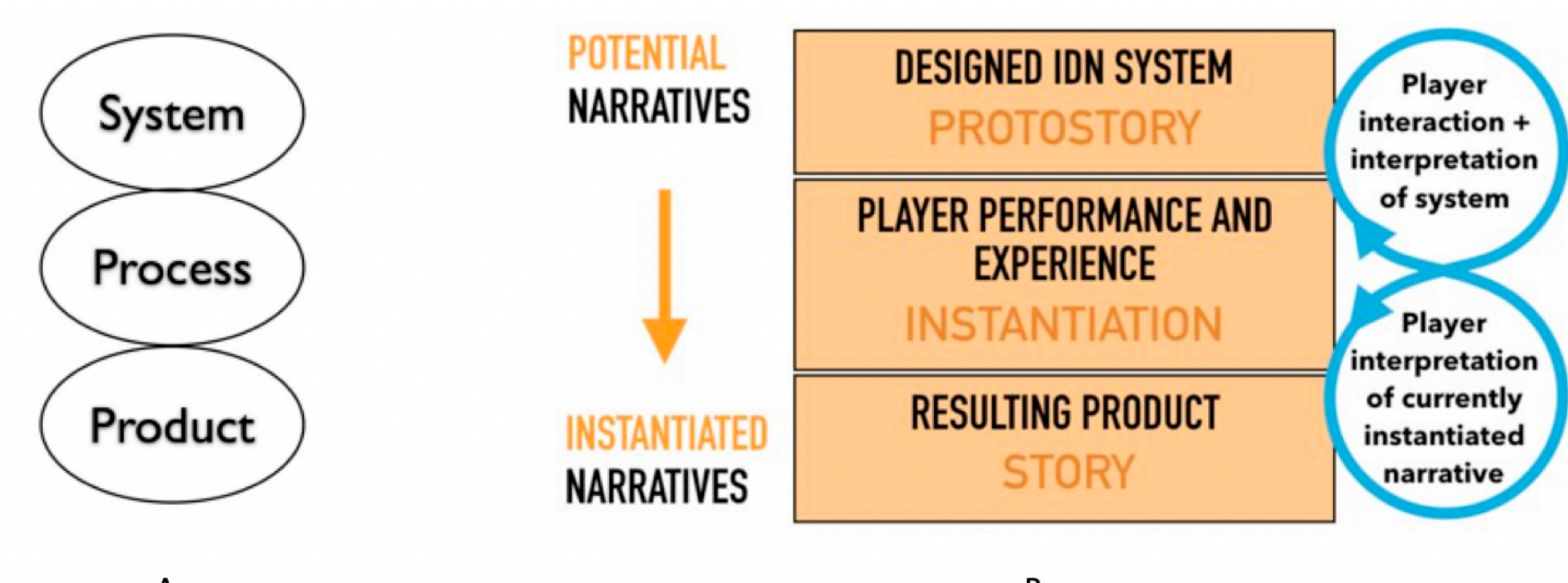

**Figure 2** Koenitz’ SPP model, where A outlines the relation between the 3 sup-parts of the IDN model, and B demonstrates how the three parts interconnect from the point of view of the interactor/player (adapted from [117] for A and [118] for B)

IDN Authoring = Design

Having outlined the “what” of IDN design, authoring can be considered as a complex endeavour that addresses content selection, mode of interaction, audience perception, and narrative generation [119]. So, how should it be done?
As defined in [120], there are essentially a maximum of 9 canonical processes[25] an author passes during production. The processes cover:

- Premeditate, the initial ideas about media production are established.
- Construct message, where an author specifies the message(s) to be convey
- Query, where a user retrieves a set of process artifacts.
- Create media asset, where media assets are captured, generated or transformed.
- Annotate, where annotations are associated with media assets.
- Organise, where process artifacts are organised according to the message.
- Package, where process artifacts are logically and physically packed.
- Publish, where final content and user interface is created.
- Distribute, where final interaction between end-users and produced media occurs

Figure 3 outlines the relations between them.

---

of computers, smartphones, augmented realit (AR), mixed reality (MR), virtual reality (VR). [115]. Existing applications that can lead to this vision of avatar driven socialising or be considered part of the metaverse are Second Life (https://secondlife.com/), work spaces such as Gather Town (https://www.gather.town/), or popular games including Habbo Hotel (https://www.habbo.com/), World of Warcraft, Fortnite (for those development engines such Roblox (https://corp.roblox.com/), Unity (https://unity.com/), Blender (https://upbge.org/#/) or Unreal (https://www.unrealengine.com/en-US) need to be considered).

[25] According to [121], a 4-process circle is sufficient (paper phase or idea treatment, prototype, production, testing), whereas [122] breaks the process down to 3 conceptual steps (idea generation, implementation, simulation).

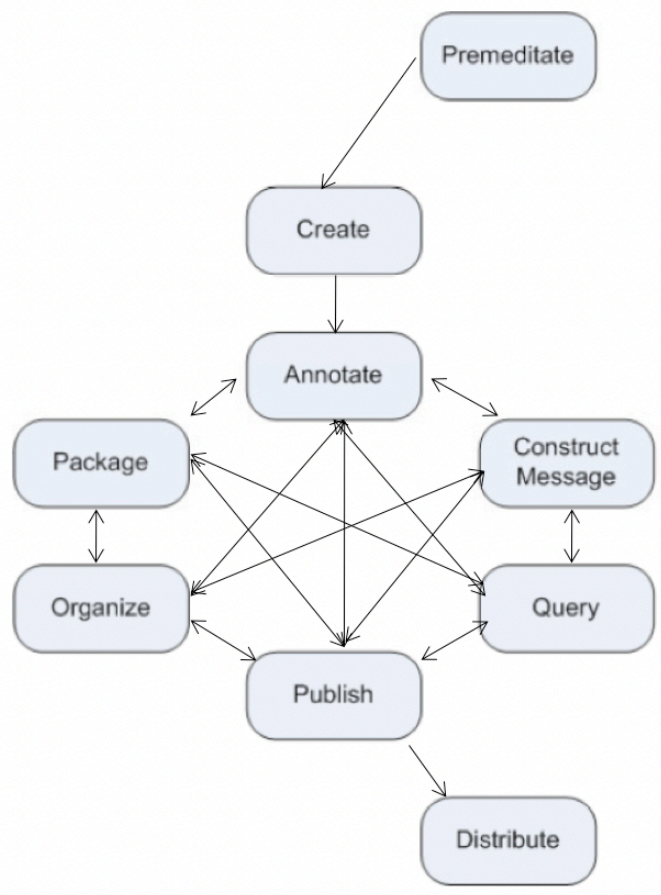

**Figure 3**: Canonical processes in authoring of a media artefact.

The canonical processes can be applied in the plan-driven design paradigm (as indicated by the premediate process) but also, as specified through the interlink nature of the representation, within the action-centric design paradigm. However, even though the authoring process here is understood as a cyclic process, it essentially refers to the classical communication view on authoring as outlined in Figure 4.

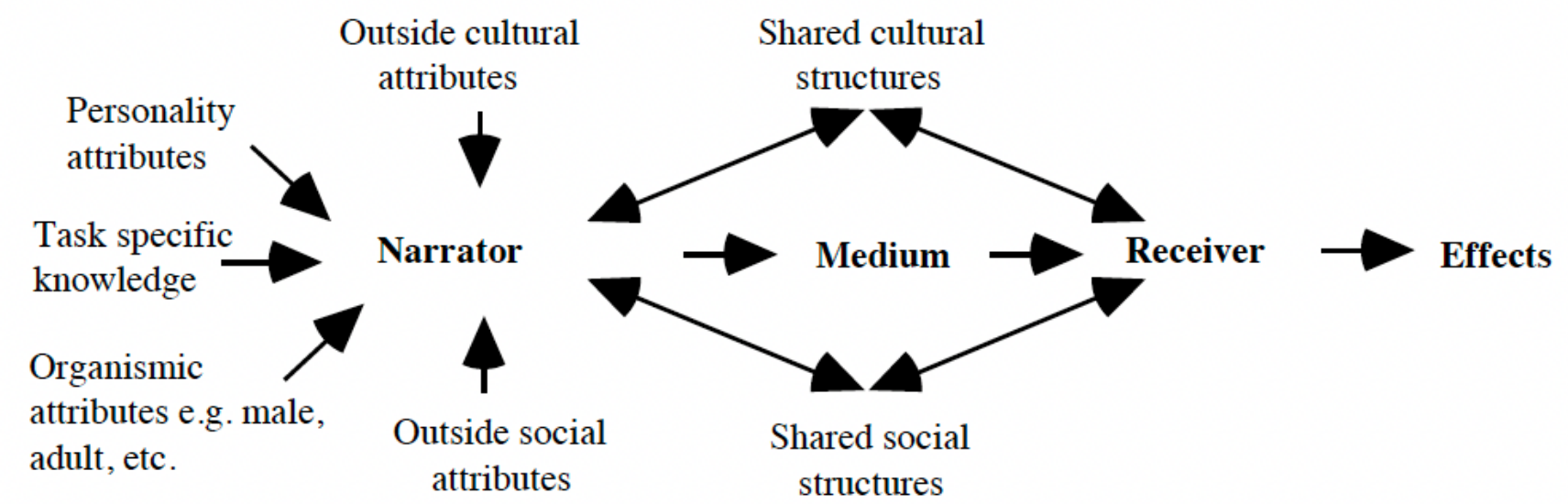

**Figure 4** The structure of communication (based on Tudor [123, p. 31)

An IDN is a communicational system that develops a representational structure of content and experience. In addition, the IDN has to facilitate free interaction, which require the design of structures, where core operational processes as observation, selection, and realisation result in meaning generation, that is established over a temporal duration in which the system adapts to interactor expectations as the major means to generate immersion. This temporality can be essentially designed in two ways, namely like a helix or a mosaic. The helix approach has been introduced in communication studies by Dance [124] and has been applied for the context of narrative by Hayles[26] [125]. An adaption of this view has been outlined in [126]. Though the spiral model covers temporality, as it depends upon the past, which informs the present and the future, it is critical as it is not structurally organised, and variables cannot be identified separately. The mosaic model[27] [127] is based on the idea of tracing various elements of a message, where the items can be separated by gaps in time, others by gaps in modes of presentation, or in social situations. This model is aligned to memory models as outlined earlier regarding discourse, conceptual dependencies, dynamic memory and schemata, but also associates with representations, such as user, argumentation, episodic, or perception models and hence facilitates the dynamicity of IDN interaction between the creator, the content, the audience and the situation. The advantage of a mosaic view on IDN authoring is that an integration with both the Canonical and the SPP model is possible.
However, the temporality of the narrative design is still problematic if one looks at the development and distribution of content. Ricoeur's work [16] is a reasonable start point to see how the IDN process could be aligned with particular content and related actions. This is outlined in Figure 5.

[26] "Yet even the most insightful and reflective of the cyberneticians stopped short of seeing that reflexivity could do more than turn back on itself to create autopoietic systems that continually produce and reproduce their organization. Heinz von Foerster's classic work Observing Systems shows him coming to the threshold of a crucial insight and yet not quite grasping it: the realization that reflexivity could become a spiral rather than a circle, resulting in dynamic hierarchies of emergent behaviors." [125, p. 242].
[27] Which aligns with Murrays idea of agency and transformation.

Mimesis 1

Composition

Structure
Conceptual network (action)
Agent, Goal, Themes, Motives, Conflict, Interaction (cooperation, competition, comparison), Results

Paradicmatic order of action => syntactmatic order of narrative what, why, who, how, with whom, against whom

Symbolic resources
Signs, rules, norms, ethics

Temporal
Present, Past, Future
Story time
Real time
Actual time (reading , reflecting)

Mimesis 2

Perception

Guided by Expectation and conclusion

Plot 1
Mediation between the individual events and the story at a whole

Plot 2
Emplotment => bringing together the heterogeneous as agents, goals, means, interactions, expected and unexpected results => episodic and configurational dimension

Plot 3
Generalization and synthesis

Mimesis 3

Application

Configuration, refiguration, reading, and reflection

Marks the intersection of the world of the narrative and the world of the interactor.

**Figure 5** Adapted from Ricoeur's theory of a correlation between the activity of narrating a story and the Temporal character of human narrative experience ([16], Vol. 1, pp. 52 - 90

Though this overall cyclic framework is rational because it is oriented towards the understanding of narrative as a hermeneutic process[28] that acknowledges the role of the perceiver in mimesis 2 and 3, it is still not sufficient to explain the how, as it cannot cover the system aspect, which is crucial for the interactor's double hermeneutic process as described in the SPP model.

Taking all the above into consideration it seems sensible to approach the design of IDNs finally from the point of view of the human who does it. Murray calls this person the cyberbard [87, pp. 208-213], a composer and declaimer of epic verses. The composer reveals and shares views but the declaimer through the telling of the story becomes psychically close, develops a connection to the audience through the communal experience, and through the responses of the audience a well-trained bard moulds the tale according to the needs of the audience and/or the location or environment of the telling. IDN authoring is not understood as a classic Aristotelian approach to narrative generation but rather as a fragmented, distributed and interactive process that still results in a coherent presentation that potentially spreads across media but addresses, and hypothetically even predicts, the audience's potential intellectual and emotional state at any time. The composer of an IDN is a source collector, the designer of telling (but not the teller), the provider of motivated exploration, reflection and experiences over time, as well as an expectation engineer. Seeing it this way the term cyberbarde might not be the right term, as it encompasses telling. Narrative designer would be a good description had it not already been conquered by contemporary video game development, where it describes the design of the narrative elements of a game based on how players interact with its story[29]. Creator is used often in the IDN community but in this article the term "narrative engineer" is from now on used, as it addresses the technical side of the authoring process and keeps no doubt that all he or she can provide is the means of telling without doing it. Figure 6 outlines the steps the narrative engineer has to perform and how this design thinking procedure relates to the double hermeneutic loop of the audience (a large number of individual interactors) as outlined in Koenitz' SPP model.

Looking at the blue part of Figure 6, it is obvious that IDN design is essentially human-centred in its design goals, as it addresses human needs, competences (cognitive, knowledge, and skills) and behaviour but it also requires that the environment in which the engineering has to be performed is designed in similar terms. Looking in more detail at the processes to be performed it seems that IDN design leans towards the action-centric paradigm as the predictive nature of the development asks the narrative engineer to simultaneously refine the mental image of the design object (the system) based on the actual perception of the context in and for which it needs to be designed (process and product).

---

[28] Hermeneutics is the theory and methodology of general interpretation, including interpretative principles for understanding and communication (including written, verbal, and non-verbal) though the emphasis lies on the wording and grammar of text [128,129]. Understanding a text hermeneutically is considered in form of a cyclic process, in the way that one's understanding of the text as a whole is established by reference to the individual parts and one's understanding of each individual part by reference to the whole [130 - 131].

[29] Molly Maloney and Eric Stirpe describe the role of a narrative designer in contrast to the game designer in their talk at the 2018 Game Developers Conference. See https://www.youtube.com/watch?v=8FgBctI5ulU

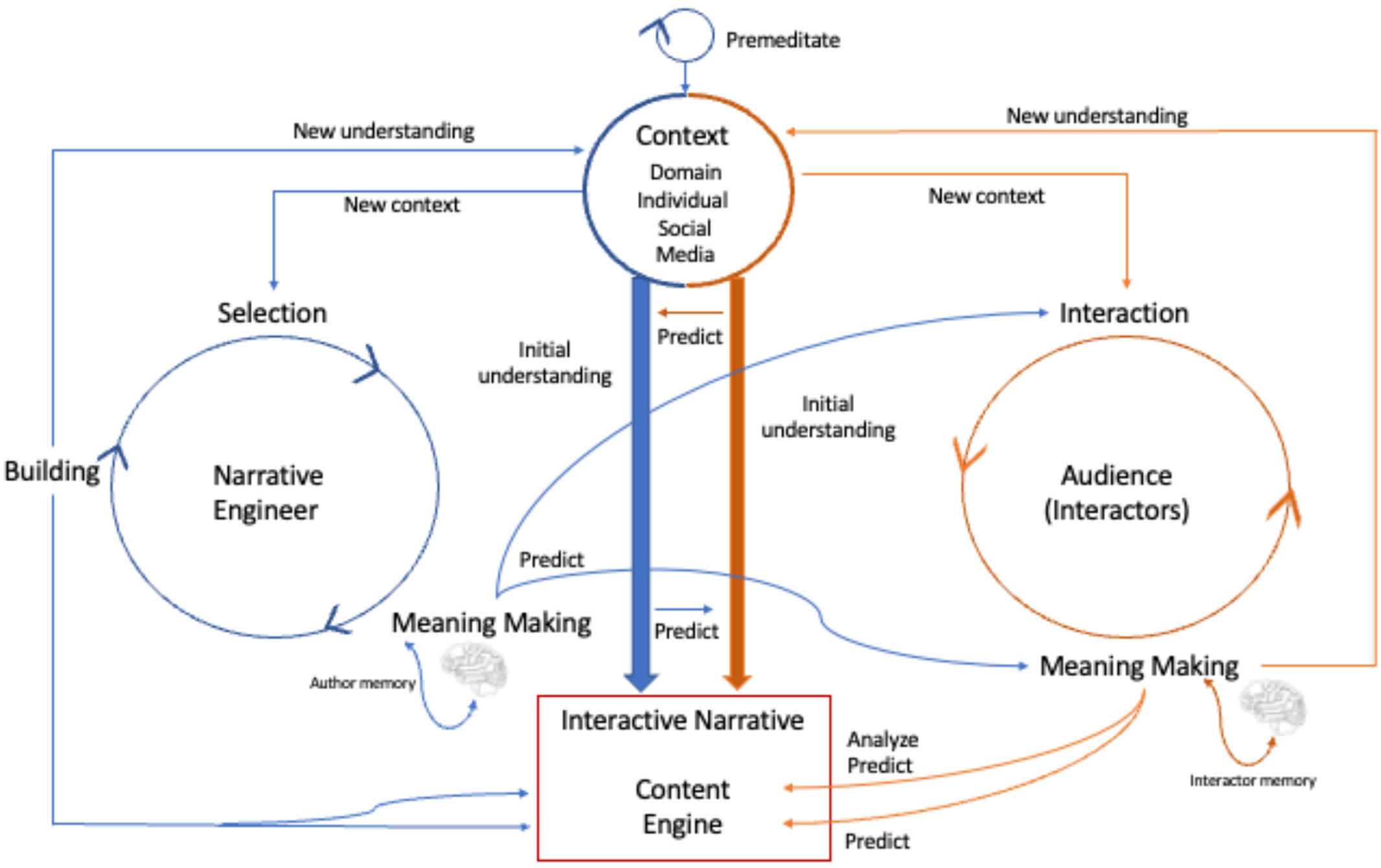

**Figure 6** Hermeneutic dynamics between authoring and perception of an interactive narrative

This does not mean that a rational, plan-driven model can be excluded, as the engineering of the system requires discrete development sequences, only that the traditional view on those (i.e. pre-production, production, and post production) and related tasks is not applicable on the whole design process. IDN authoring needs to be considered as a creative human process establishes potential narratives, where the narrative engineer imagines the story in multiple iterations, as well as multiple modes, exploring many possible choices and outcomes and can make those possibilities available to the system so that the audience can collaborate with the system to achieve narrative flow in a personal way. As the formal categories of IDNs can differ, the skill sets of a narrative engineer might lean towards teller or builder, with a favour for a rational or action-based approach for working, it is not obvious how IDN design then can be supported. This point still needs to be addressed.

Tools

As IDNs are about digital artefacts, the narrative engineering process necessitates digital tools[30]. Such academic or community tools are widely available, either running on a pc, web, or cloud infrastructure. Good review papers are [132 – 135], where in particular [133] is relevant as it provides a categorisation and description framework for IDN authoring tools (9 categories and 38 descriptors for tool analysis and comparison) based on which around 300 tools have been surveyed and classified[31].
The richness of the available tools is an indicator that practitioners as well as researchers still see the need to develop tools that cover their understanding of what an IDN needs to address and convey, and in which type of communication this should be done. An example of this line of thinking is the development of the web documentary “Pregoneros or Medellin”[32], which has been conceptualised from the beginning (ideation) as the realisation of a virtual street walk. The story was envisioned as a stroll through the street and its urban soundscape, where the street-vendors shouts played an important role and should be experienced as the invitation to meet them. The idea of the story is not so much a consistent narrative but rather an experience of a documentary space built out of interconnected stories. The team describes in much detail how existing web technologies have been utilised in combination with newly invented capture technologies to finally create the IDN[33]. In this example, the teller paired with a particular type of engineer (web) to provide the means for an envisioned experience. Other tools, as

[30] Tools that are related to the generation of particular media units, to be used in an IDN as a source, such as word processors, image, video or audio tools, or game or world design engines will not be covered in this text.
[31] See also IDN Authoring Tools Resource. interactivenarrativedesign.org/authoringtools/appendix.pdf.
Last accessed 15.05.2022
[32] https://pregonerosdemedellin.com/#en
[33] Thibault Durand (2015). How we created an immersive Street Walk Experience with a GoPro and Javascript
Sharing back the making of the web-documentary Pregoneros de Medellín (Colombia). https://medium.com/@tibbb/how-we-created-an-immersive-street-walk-experience-with-a-gopro-and-javascript-f442cf8aa2dd last accessed: 12.05.2022 outlined in literature, also tend to address the technical side but furthermore try to provide an easy access, as they are basically developed to help non-programmers to build IDN systems[34]. Their design aims at the development of open navigation but their user interfaces and underlying narrative technology in form of templates, modelling, analytics, and rendering mechanisms, are oriented toward the sense-making affordances (syntax and semantics) of the main medium they operate on, most often text or visuals (graphics or video). In complexity, they alter between tools that facilitate the integration and organisation of recorded content up to fully generative, code-based tools (parsing, graphic rendering, etc), often applied in the domain of narratives in games. This does not mean that generative tools necessarily require more advanced technical skills because manipulation for personalisation of the recorded material requires the mastery of technologies such as indexing, media synchronisation, or ontology handling. The simpler the media and the more reduced the explorative and adaptive capabilities of the envisioned end system the more amateur-friendly the tool.
As the tools are closed environments an exchange of information or procedural components is nearly impossible. This leads to the situation that still today authors interested in interactive narrative have to look for the right tool to support them, which becomes increasingly frustrating, if an author changes the expressive space, which asks to learn more new tools.

## Vision

Considering the still existing problems in IDN authoring, the remaining text outlines a potential environment for supporting a narrative engineer during the creation of an IDN. The central design paradigm is the facilitation of a collaboration[35] between itself and the narrative engineer. This manifests essentially in the way how the environment adapts to the narrative engineer's skills and ways of working as well as to the fluidity of how the sequences of analysis, design and implementation are undistinguishably connected. Hence the suggested environment follows an action-centric design approach, where the relevant design actions are supported by the environment through intelligently performing services[36].
The overall structure of the envisioned environment follows Koenitz's SPP model (see Figure 2), where the System part (S) is the one that mainly needs to be addressed by the environment. The Process part (the first P) is used to validate the design (following the action-centric paradigm). The main representation of the S and P part adapts Ricoeur's Mimesis model (see Figure 5[37]), resulting in a 4-process authoring model[38], covering Ideation, Meaning Making, Interaction, and Validation. Figure 7 provides a sketch of the potential architecture.
The ontological base is Chatman's idea that a narrative is organised as a representational system based on surface structures (expression) and deep structures (content), where each of those distinguish between substance and form (see Figure 1). The environment focuses on the design of deep structures, where the authoring of the surface structure is mainly left over to tools that handle media creation (i.e. image, video, 3D, and audio tools[39]). The underlying representation structures regarding the meaning of the protostory space are designed in form of knowledge graphs[40], as they best represent ontological spaces.

---

[34] See authoring tools for hypertex [136 - 138], natural language oriented environments [139 -140], drama and role-playing authoring (emerging narratives) [141-148], visual media related authoring [149-152], media agnostic tools [153], socially aware multimedia authoring [154 - 157], web-based authoring tools [158]. With resepct to [152] see also https://teenage.engineering/now#B-1

[35] This idea is not new. See for example [159 - 162].

[36] Services here are considered as systems (mainly Generative AI tools) in the context of Luhman's communication theory. It outlines that participation in communication requires that one must be able to render one's thoughts and perceptions into elements of communication. This can only occur as a communicative operation (thoughts and perceptions cannot be directly transmitted) and must therefore satisfy internal system conditions that are specific to communication: intelligibility, reaching an addressee and gaining acceptance (the encoding of information in the system reduces complexity of the environment). Each system works strictly according to its very own code and can observe other systems only by applying its code to their operations. (see [163], p.185 and Chapter 6).

[37] In this article, we cover only Mimesis 1 and 2. The impact of Mimesis 3 on this article is outlined in the validation and conclusion section.

[38] In terminology, we follow the INDCOR IDN taxonomy [164].

[39] Consider tools for text (any editor), 2D graphics (Illustrator, Photoshop, Indesign, Photo, GIMP, DALL-e 2, Midjourney, Firefly), video (i-Movie, Final Cut Pro, Adobe Premiere Pro, Lightworks, Sora, Gen-3 Alpha), Audio (GarageBand, Steinberg Cubase, FL Studio), Games (Unity, Unreal).

[40] In the context of knowledge representation and reasoning, a knowledge graph utilises a directed graph-structure (nodes and relations between them) to store descriptions of entities (i.e. objects, events, situations or concepts) in a way that semantics or relationships underlying these entities are encoded [165]. The ability to reason about the knowledge space established by the graph is achieved via inferred ontologies in combination with a reasoner. Relevant papers that demonstrate the use of knowledge graphs in the context of IDN authoring and utilisation are [166 – 169].

The authoring environment is envisioned to be web-based, within a cloud-based environment (the interface is the client, the storage and analysis, and actual system is server-based). The authoring environment is open-source, as the different sub-systems addressing content, audience and media structures should be developed by experts for the articulation and analysis processes and template modelling for the different types of IDNs.
As we investigate the environment in the context of a collaborative effort, where the authoring system mainly supports based on knowledge about the narrative engineer's ways of working, the potentially needed content and expression structures and related ontological representations and narrative cognitive models (for fore- and backend), we apply the approach of hermeneutic phenomenology, resulting in dynamics as outlined in Figure 6. As the authoring environment is a system that is able to finally generate the IDN environment, including the frontend and backend engines, the applicability of the designed hermeneutics supports the authors in a way so that the final mapping of anticipated behaviour and actually performed interaction and meaning making are close.

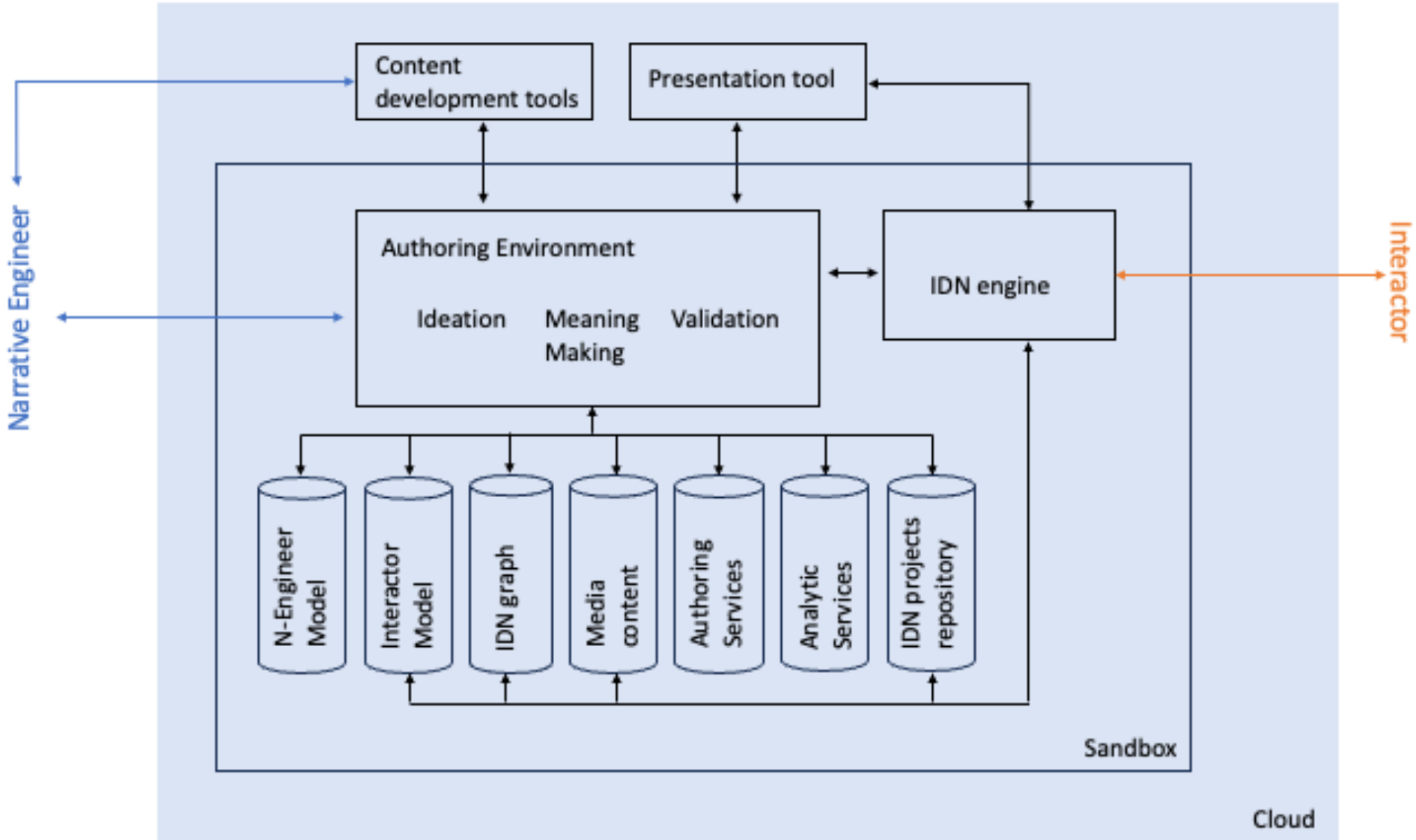

**Figure 7** Authoring environment architecture

That is why the process part at some stage of development needs to be integrated as a validation step[41]. The adaptation is based on the system's memory of the engineer's skills and preferences shown in previous IDN designs. This way of realising the authoring process as action-centric design requires that the environment knows where and on what the narrative engineer is active with. Therefore, the overall design of the interface separates the different phases of the authoring process but facilitates the easy move between them as seems necessary. Examples closest to what can be imagined are presented in Figure 8 (a and b).

[41] Technically speaking this validation is based on a simulation (created by the narrative engineer), unless the engineer incorporates potential users in the testing.

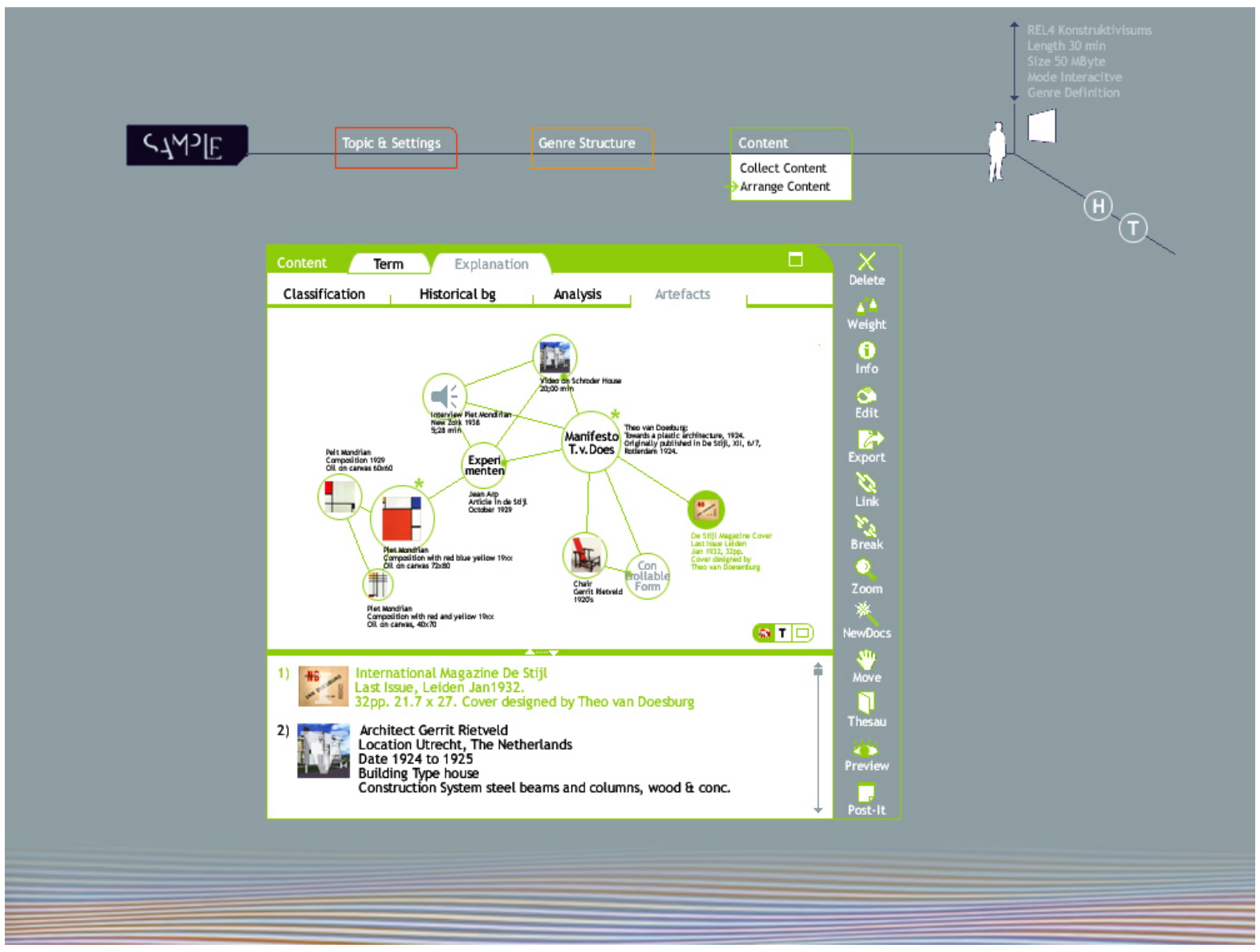

a

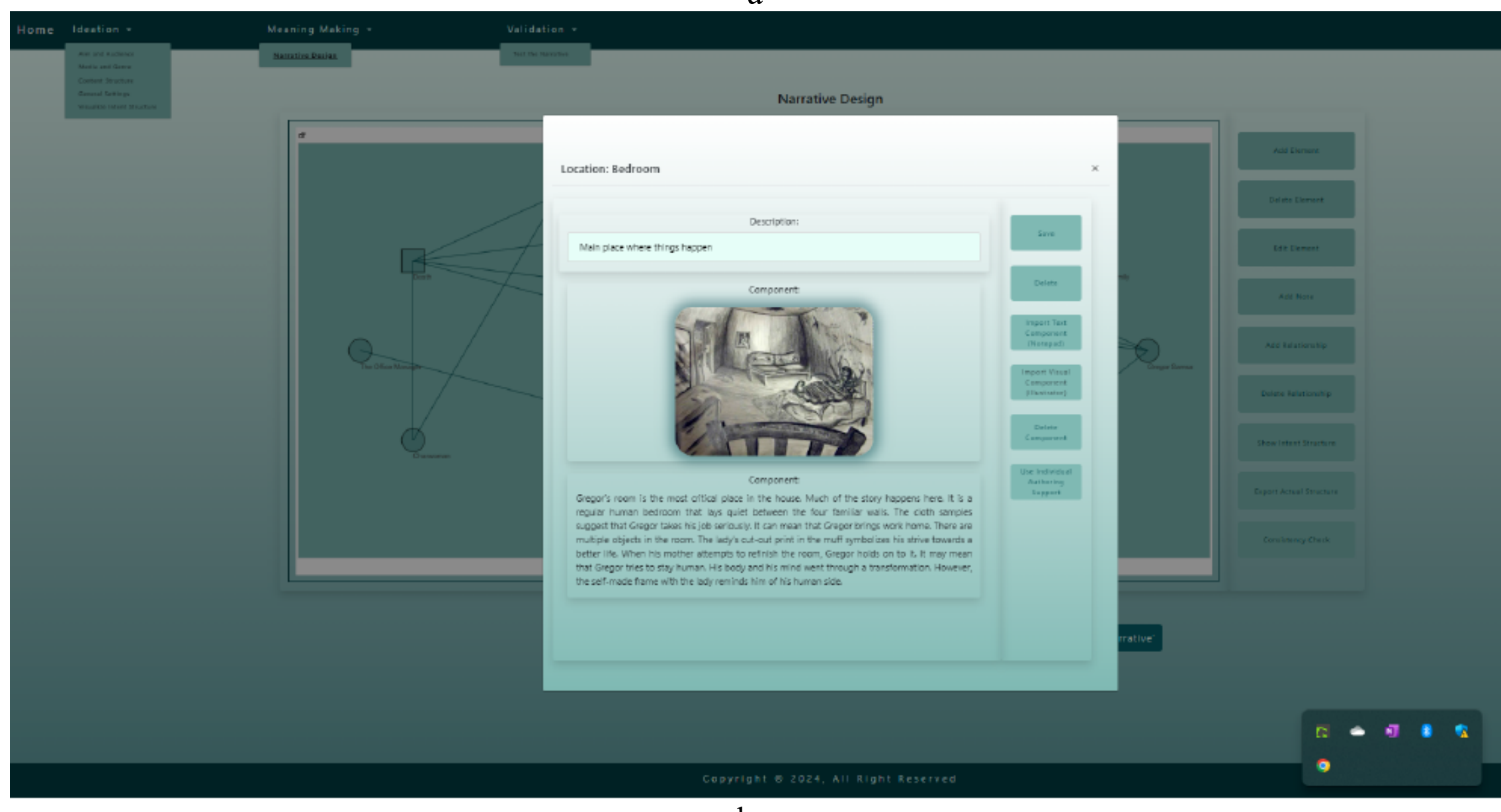

b

**Figure 8** a) The SampLe multimedia presentation authoring interface, based on underlying knowledge-graph structure that contextualises representations of discourse, genre, and visual media [151] and b) Sandbox prototype showing Meaning Making - Narrative Design - Edit Location Element (Fiction Media, Text+Visual Modality) [170].

The proposed authoring model foresees a narrative engineer for whom the basis of authoring is to support comprehension of content to achieve meaning making related to the topic under investigation. The aim is to suggest an environment in which a narrative engineer can create controlled agency that is guiding rather than directing but follows persuasion rather than poetics. Figure 9 and 10 graphically represent the various components we consider necessary in such an environment. Figure 9 represents the view on representation models the author can make use of while generating the proto-narrative space, whereas Fig. 10 represents the resulting forms of interaction and related processes. Finally, the outlined vision is considered distributed, meaning that several authors can develop narrative spaces with the same engine while the finished works can be accessed by a limitless number of interactors (with the individual experiences adapted to the required level of comprehension).

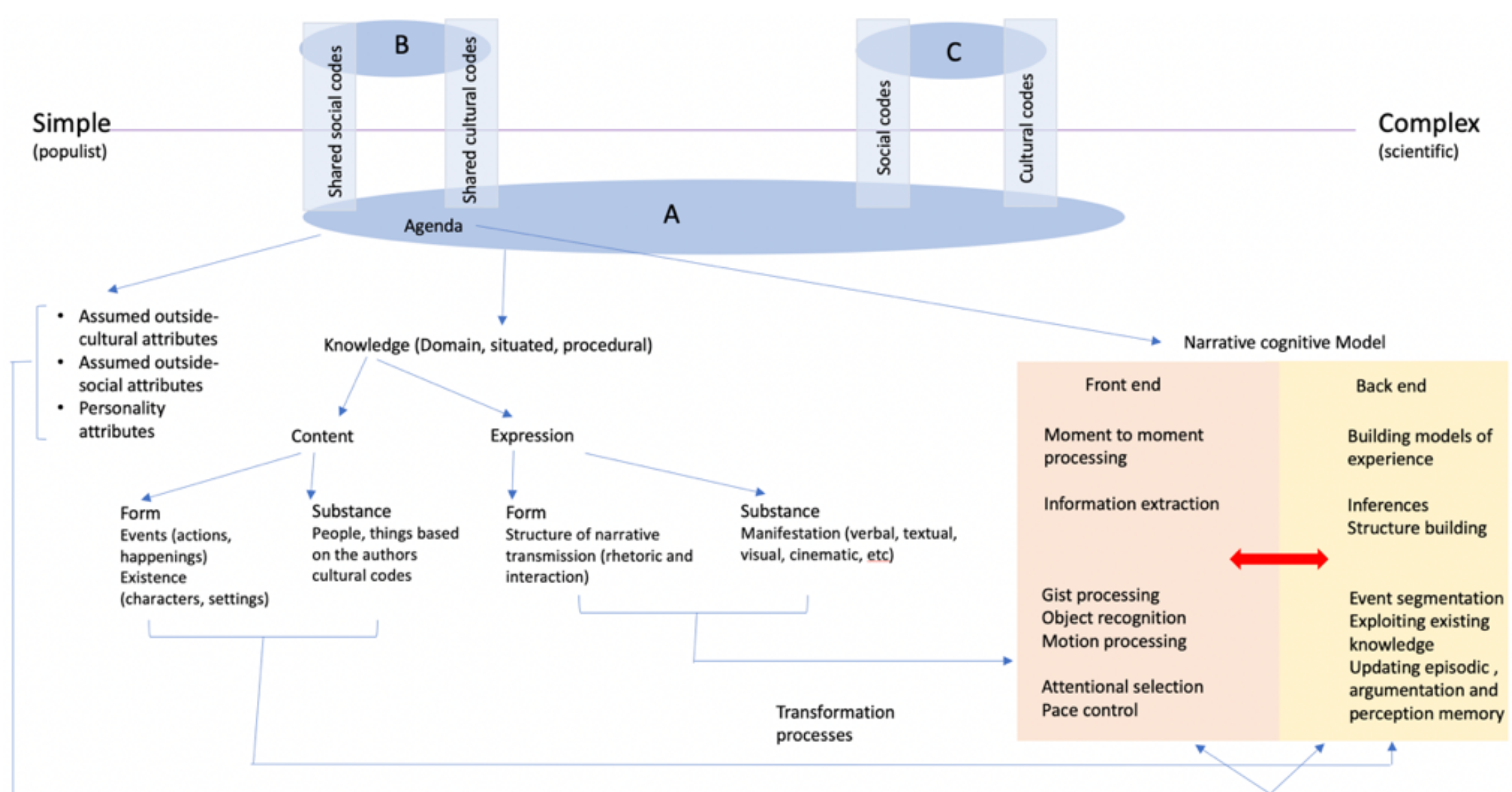

**Figure 9** Knowledge representation models of the authoring environment used during authoring (purple represents the interface between the narrative environment and the interactor, blue ellipses represent the narrative engineer (A) and potential interactors (B and C), blue arrows represent detail relations of knowledge necessary for IDN design and building).

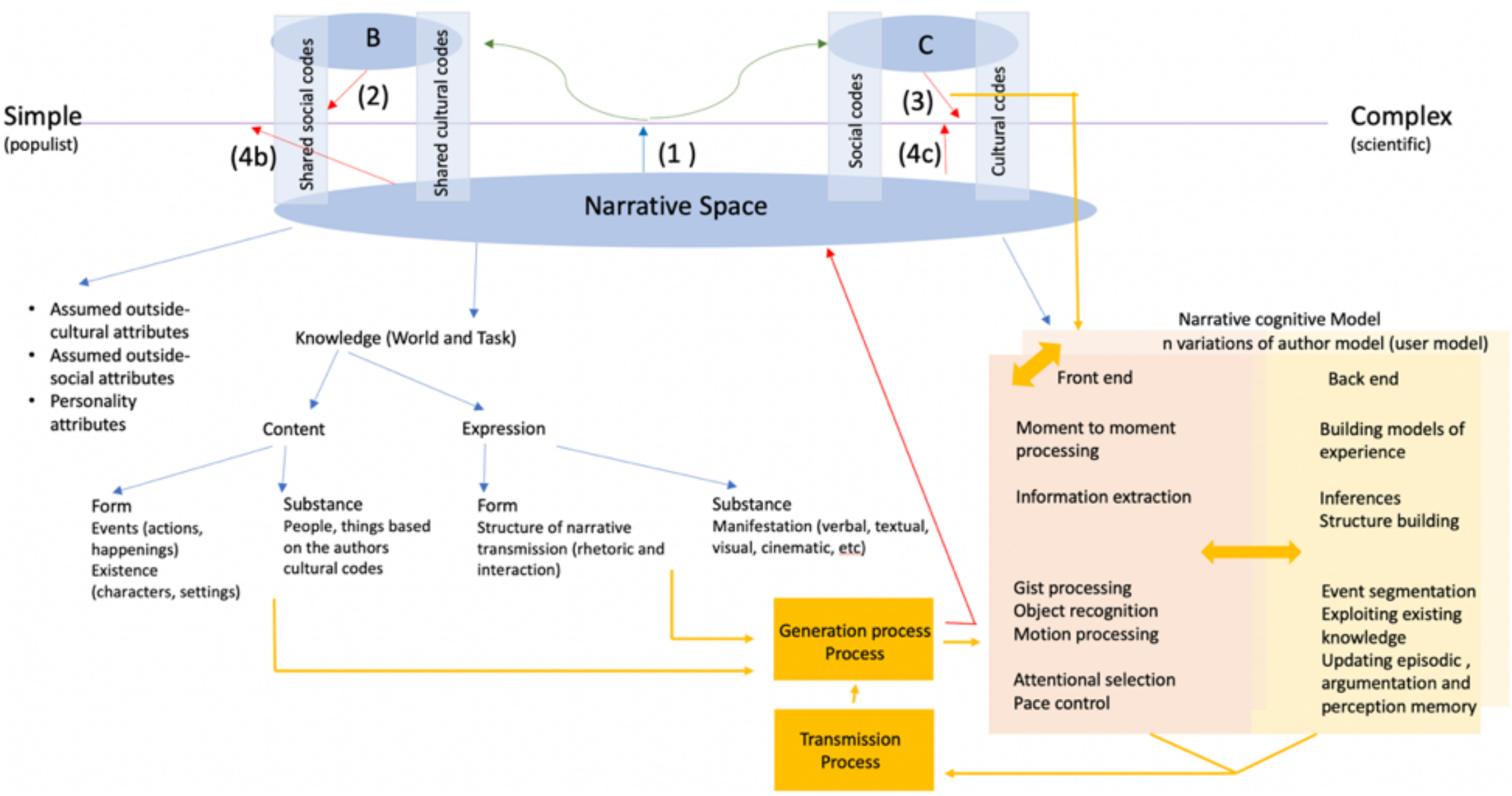

**Figure 10** Process space of the authoring environment when the narrative is explored (purple represents the interface between the narrative environment and the interactor, blue ellipses represent the narrative environment (S) and potential interactors (B and C), blue arrows represent detail relations of knowledge represented in the system, and orange outlines the adaptation systems of the system towards the comprehension needs and meaning making processes of the interactors).

Having provided the structural background, we can now outline the 4 authoring processes. This is done based on a scenario, where a journalist is taking the role of the narrative engineer, with the aim to establish an argumentation-based multi-media web environment synchronised with its newspaper online environment, in which interactors can compare the different views of members involved in the Sudanese civil[42] war (2013 – ongoing).

---

[42] https://en.wikipedia.org/wiki/Sudanese_civil_war_(2023-present)

*Ideation*

Before the journalists start travelling, he sets up a plan to establish what type of material needs to be collected. The system supports this by asking questions about the core ideas regarding the communication goal, the key audience, and the key concepts of the protostory space. The way how the plan is developed follows a guided discourse structure, guides from the most general towards the level of detail the author wishes to achieve[43]. For being able to do this the system makes use of established knowledge structures [44] and already available knowledge about the journalists work environments and mannerism and preferences[45]. The conceptual elements cover:

| | |
|---|---|
| Type | Non-fiction (biography, journalism, transcript poetry and historiography); historical-fiction (anecdote, myth, legend, history), fiction (prose, short story, novels, narrative poems), performance (games, live or recorded performances). |
| Audience | covering age range, educational level, conceptual background, wants, needs, intentions |
| Complexity | Content (simple, complicated, complex), Structure (simple, complicated, complex) |
| Media | mono, visual, audio, mix ….. monochrome, colour |

Based on those basic structures the authoring environment then can, if whished for by the interactor investigate deeper on the level of content and expression structures, such as

| | |
|---|---|
| Content | Agents -> what, why, who, with whom, against who, actions |
| | Objects |
| | References (cultural, economic, ethics, etc) |
| Expression | Argumentation types (i.e. deductive, inductive, argumentation/rhetoric schema[46]) |
| | Rhetoric (e.g. use of i.e. symbol, trope, metaphor or metonym) |
| | Style (exposition, retardation, digression, omission, redundancy) |
| Notes | Free collection of ideas.[47] |

For the journalists this means that before visiting the Sudan he can already sketch down the potential type of work (interactive journalistic exploration space), the audience (template from previous works in this tool), the expected complexity level (complex), that he will collect data in form of images, videos, and text, and he can already establish the relevant people, organisations, regions in Sudan and related regions, the role they play in the conflict and the relationships between them.

The provided information is used by the authoring environment in two ways. First, with the outlined information the authoring environment makes available the relevant representation and generation units, based on the chosen conceptual elements. It also provides links to already established IDNs by the actual creator(s) or establishes connections to related works in form of services, such as detecting missing or isolated content elements, or selecting of or checking on expression aspects. In our case here the journalist might have shown a preference for the Toulmin argumentation model and so the system will provide metadata structures to facilitate the annotation of media items based on their role in an argument (between agents but also conflict lines), including the preferred LLMs. It also provides the necessary tools for the design of the interaction engine, such as functions or libraries to analyse graph structures, or the associated generalised models for the front and back ends (see Figure 9). Second, the established plan is already transformed into a meta-IDN representation (the knowledge graph representing the communicational goals, the core concepts and relations between them, against which the system can in the meaning making process compare if the aimed for IDN is still under development.

Once done, the environment makes accessible the remaining process units of the system (meaning making, interaction, validation) so that the journalists can start any time to work on the actual IDN.

---

[43] Dramatica (https://dramatica.com/) provides an example of a plan-based story development tool for the development of film, where [170] outlines an approach that is closer to what is presented here.

[44] Already existing narrative taxonomies and ontologies [66, 164, 171 - 176]. Other relevant sources are Dpedia (https://www.dbpedia.org/), Wikipedia (https://en.wikipedia.org/wiki/Wikipedia:Contents), The Living Handbook of Naratology (https://www-archiv.fdm.uni-hamburg.de/lhn/contents.html), Stanford Encyclopaedia of Philosophy (https://plato.stanford.edu/entries/fiction/), Plotto (https://github.com/garykac/plotto). Other potentially relevant vocabularies are the Getty thesauri AAT, TGN, CONA, ULAN, IA, CDWA (https://www.getty.edu/research/tools/vocabularies/).

[45] If the system is used for the first time it runs an instantiation discourse to collect information for example on preferences on media editing software, potential publication environments, etc.

[46] Relevant schemata are for example the Toulmin model [177], which has been used in an interactive Vox Populi [152], or the Periodic Table of Arguments (PTA) in combination with the Argument Type Identification Procedure (ATIP) [178, 179].

[47] [180] outlines how the use of GenAI (here GPT-4) can facilitate support in this creative process. See also [181] that outlines for 15. authoring tasks how GenAI can be usefull (here in particular LLMs, where in particular Claude (https://www.anthropic.com/claude) seems promising). For more graphical applications see [182, 183].

*Meaning Making*

The meaning making phase addresses the actual building of an IDN. As already pointed out, once the narrative engineer starts working on his or her project, the relevant services, underlying knowledge structure concepts, potentially necessary storage structures with related analysis methods, and links to the preferred content authoring tools are already made available in form of interface menus.

The design flow within "Meaning Making" begins with specificity. Material collection is considered specific, as the narrative engineer works on single information source (whatever medium) at a time. This phase supports search for material or its creation with the chosen creative tool (i.e. a word processor) but also establishes its potential role and meaning within the chosen aim, audience, content, interaction context. Looking at our journalist, this means that once content has been generated, i.e. audio or video files with interviews, images or videos representing location descriptions, graphs the represent socio-economic or texts that facilitate analytical descriptions of political or historic contexts, the relevant files are stored in the systems content repository for this this particular IDN project. In the meaning making interface part such a file appears as a node, which already contains the link to the related file and the media type. The engineer can now name the node, assign in the provided node character, location, object or information label, and define the argumentation role this node supposed to take. For Toulmin this might be Fact, Warrant, Backing, Conclusion, Rebuttal [177]. If only a fraction of a temporal file is relevant (i.e. the narrative engineer decided to not split video or audio files in smaller junks), the knowledge engineer can provide start and end pointers to the address of the file (see [156, 184]). In case the role alters over time, such changes can be performed any time. The system then checks for graph consistency and suggests related changes. In case new material is required that does not exist yet this can also be provided any time. The journalist can start linking nodes together, making use of the established relation types of the chosen ontology. Thus, over time a large knowledge graph emerges.

This knowledge graph is stored in the assigned IDN graph repository (see Figure 7). This graph represents the current state of the designed protostory but also contains the complete design and editing history[48], i.e. all alterations regarding notes and relations, node insertions and deletions (based on time stamps in nodes and relation descriptions). How such a file can be generated during the meaning making process is outlined in [169]. This network can be made visible in the structure part of the simulation tool in two representation forms, where the narrative engineer can zoom in and out particular contextual structures. A higher-level visualisation represents the protostory where the media units display the nodes and the relations relevant feature connections. This is necessary so that the narrative engineer can check the protostory space that he or she generates. The second form of representation is a set of tables to facilitate an observation of the relations between events and concepts. In this way the narrative engineer can control the distribution of material on a level of potential reflection and comprehension[49]. This is considered the backend of the narrative space (i.e. the semantic and episodic memory).

The graph also forms the basis for the support of the system with respect to consistency and completeness established by the comparison of the graph analytics against the outlined conceptual structure of the ideation plan. Consistency here means that the system can analyse syntax (with respect to complexity) or semantics (the use of rhetoric and expression means as outlined in the plan) to provide insights if the environment addresses the potential audience as indicated. Example services already established are described in [161, 162, 169, 180, 186, 187]. The system can also check if established arguments are still open or if nodes are isolated. The support on this level is based in form of discourse where the system suggests, where the suggestions are either based on the current space the narrative engineer works in, or on general analytics that show, based on thresholds set by the narrative engineer, that mainly point to differences between plan and current environment. The narrative engineer can alter the plan (action centric paradigm). Those alterations are also noted in the graph's history representation.

In parallel the narrative engineer develops the set of processing rules that facilitate the interaction once the explanation/exploration environment is used. This unit adapts the SPECT framework [69], only that here it is not used for the analysis of narrative comprehension but rather as the means to generate for potential comprehension[50]. Our journalist designs and develops the traversal of the graph in form of interaction patterns. It is also in this part that the narrative engineer establishes role dependencies, comparison parameters and constraint sets. For our example of the journalistic argumentation space, the engineer relates rhetoric and argument concepts, such as claim, ground, warrant, backing, rebuttal, qualifier (associated in form of metadata in the relevant nodes, see above) to transmission processes based on the level of expected complexity, such as equivalence (use material as is), contraction (e.g. ellipses and compression), and expansion (e.g. insertion and dilation) depending on user specific instantiations. Internally those processes are applied to the different media types used in the narrative space (see [66, 152, 184] as examples for the adaptation of video, or [190] for branching comics). Those representations of transitions and interactions are used for the generation of the simulation of the IDN environment, so that the narrative engineer can test, in a limited way, if the planned for interaction works as planned (see the right side in Figure 6), In our example for the journalist the simulation is an HTML environment, that uses structures and code that the journalist has used in previous IDNs generated for the environment of the news cooperation he works for.

---

[48] The base for this design facilitates the work by Latour [188] and Pasquetto et al [189], who both point out that experts (they focus on scientist and scholars but that is also trye for designers or narrative engineers) develop deep expertise in their domain. They apply method and tools to collect, analyse, interpret and publish the material they combine into their work that later can become part of larger collections (for stories see LLMs as an example). Thus, this expertise should be captured during the production process itself, as later it is hardly possible to extract it in a retrospective fashion but it is necessary to to have access to it so that deeper analysis of the material is possible.

[49] Relevant annotation and presentation concepts can be found in the WWW Synchronized Multimedia Integration Language (SMIL) - https://www.w3.org/TR/SMIL3/

[50] However, the analytic power of this model is also used by the system to validate consistency with respect to user interaction and agency.

*Interaction*

Once the proto-narrative has been established and the interaction/reasoning engine implemented, the exploration of the narrative space can be made open for the public. This part of the system integrates Hardman's canonical processes namely query, construct message, publish, and distribute. Note, once the space has been made public the narrative engineer can no longer influence the narrative space but can monitor how it is used. Any cases of alteration the narrative engineer makes will necessarily result in a new version of the narrative space, resulting in essentially different types of products as outlined in Koenitz' SPP model..

As the established IDN environment is distributed, for each interactor an individual narrative cognitive model will be generated (see user model in Figure 9, which has the same structure as the proto-narrative but will be necessarily instantiated differently). The entry point for all interactors is based on the interaction model defined by the author during the meaning-making stage. Every interaction is considered as a stimulus (see 2 and 3 in Figure 10).

The IDN picks the source/query and processes its front-end and back-end parameters. Those are used to investigate the degree of complexity the interactor potentially accepts (for example the distance between selected source and original source taken from the proto-narrative representation – this depends on what the narrative engineer has defined), which support the instantiation (at the beginning) or the further construction of the interactor's narrative cognitive model. For example, social codes associated with the presented source and shared with the chosen source are considered shared social codes. between the interactor and the narrative engineer (the green lines in Figure 10 represent that this is not necessarily true as it cannot be stated what happened in the interactor's mind during the selection process). If the interactor asked a question the system identifies the part under consideration in the source and the type of question and then can determine which of the stylistic devices (e.g. exposition, digression, redundancy) might be applicable in this context of an expansion. Thus, once a stimulus is provided the interaction/transmission system, established by the narrative engineer, investigates the current comprehension and meaning making state of the interactor, takes that information into the transmission process to determine the next potential output source and then generates the next exploration sources for the interactor. Every choice will result in an update of the interactor's narrative cognitive model, which now not only contains what the interactor has perceived but also which options have not been looked at[51]. In addition, the IDN can potentially establish a preference model of the interactor's complexity preferences (e.g. rhetoric forms, stylistic preferences, media preferences and contextual information (i.e. the network between those), length of argument, etc), depending on what the narrative engineer has provided. Based on the constant comparison between the proto-narrative and the individual interactor narrative-cognitive model as well as the decision processes within the transmission/generation system it is possible that the system suggests different narrative paths to different interactors (see 4b and 4c in Figure 10). The introduction of front and backend representation systems, the transmission/generation compound, and the stimulus-based comparison of established and in construction narrative-cognitive models (see the 2 red arrows in the narrative-cognitive model section of Fig 10) facilitates the double hermeneutics as described in the SPP model. Thus, the design of the IDN is based on the modelling skills and experience of the narrative engineer but facilitates, if modelling went well, the reflection on the IDN per individual interactor.

Validation

The collaborative approach between narrative engineer and the design environment as described above so far focuses on consistency checking and problem solution advice. It certainly helps to develop coherent IDNs for various communication and articulation means. It does not guarantee that the design will work as planned. What

[51] See for a similar and rather typical approach in educational environments, such as [191]

would it need to be able to achieve that? Below we speculate about this as not too much work has been provided in the field of IDN in this respect[52].
The narrative cognitive model as outlined in Figure 9 and 10 allows for an analysis of the stored interactor models against the parameter settings of the front and backend. In that way it can be established if the interaction and meaning making processes have been modelled adequately. Here it would be necessary to integrate a level which facilitates the system to reflect on the decision process and allows to provide an explanation of why material was presented, an approach which is currently named as explainable AI [195, 196].
Such methods can be used to support the narrative engineer as well the interactor of the established IDN. A simple approach with respect to the explanation on joke construction and validation can be found in [66]. Under which circumstances and in what form this information is distributed to the narrative engineer remains a problem, mainly because the published IDN cannot be altered easily as every change will influence further interaction. Moreover, the qualitative data analysis describes behaviour but not necessarily other cognitive, reflective, or experiential levels of interaction and comprehension (see here [197]) for potential cognitive measurements).
In fact, Ricoeur's 3rd level of mimesis could be used for establishing a correlation between the activity of narrating a story and human narrative experience[53]. As the actions of each individual interactor can be recorded[54], at least reconfigurations and retelling can be analysed. Achieving an understanding of reflection per interactor requires to open the system in a way that interactors can comment on their experience[55]. What this actually means for narrative flow is unclear. Certain articulation techniques facilitate ways of note taking or commenting easier than others (i.e. a documentary asks for this type of comments as reflection is an essential part of its argumentative nature, whereas in certain types of game environments the flow and hence the experience can be negatively influenced when such interruptions are integrated). As the outlined environment separates the different articulation and media concept spaces in individual systems, it is possible to further develop in this direction on a fine-grained basis. Yet, what that means for the actual authoring process is by the time of writing unclear.
Considering the case that all authoring as well as interaction data can be stored, then the suggested environment can facilitate a far larger type of validation. As the amount of generated IDNs of different forms grows, if the community agrees to work with this environment, it is then possible to gain insights in IDNs in general and the various sub-systems in particular. The advantage for the domain then is that the authoring environment turns into a research tool that allows detailed observation of relations between system, process and product and hence would greatly help to form a general theory of what the idiosyncratic concepts and processes of various forms of IDNs are. At the same time this theory can be compared against the stored interaction logs. In other words, the authoring environment that reflects the design processes of an IDN on micro and macro level, also designs and shapes the research and finally understanding of the domain itself. Yet, this requires new forms of data analytics, covering aspects of context and causality, of which the field currently only has a rather rudimentary understanding, also due to the fragmentation of work environments and hence instantiations of IDNS that are very hard to compare.

## Conclusion

This chapter showed that an interactive digital narrative must be considered as a communication system that facilitates meaning making in the head of the interactor through establishing relations between available content and interactions performed on it. From an authoring point of view the design of an IDN is a complex process where the choices made by the narrative engineer strongly influence the experiential, meaning making and reflective means for the interactor. It was also revealed that the choices are driven by experience and skill but largely also on predictions. It was argued that the current approach towards IDN authoring, namely the provision of very specified tools, leads into a fragmentation of IDN understanding and necessarily points into a dead end. It was suggested that an adaptive sandbox system should be considered as an approach to address IDN authoring, where the creative process considers goals and constraints through a rational model that feeds into a fluid action-centric paradigm that facilitates a design process, where the sequences of analysis, design and implementation are undistinguishably connected. The aim of the authoring environment is to facilitate a space for inspiration, ideation, and implementation through which the narrative engineer refines conceptual as well as implementation ideas and explores new directions until the final IDN has been established. The design of the authoring environment aims for collaboration between the environment and the narrative engineer on the level of facilitating coherence, based on an integrated system approach, where sub-systems represent the dynamics of interaction for the relevant articulation techniques through various layers of complexity in principles and procedures. The suggested approach is an invitation to see human intelligence and creativity as the basis of art, where AI can be a helping tool.
The development task will be a colossal engineering challenge, but the advantage is that over time a living and eventually reflective domain memory of concepts, skills, and established IDNs will emerge. The essential problem is not so much the system side but rather the clear understanding how the collected data can be interpreted to form a general IDN theory. Thus, while building the challenge will be the validation side of the problem.
The outlined approach might be considered as too complicated to be achievable. However, the idea is not to establish this system in one go and also not as a system for all. The subsystem approach allows to focus on particular IDNs first, i.e. documentaries or journalistic-based explorative and reflective spaces. In a way, the outlined approach facilitates the collection and integration of already established procedural and representation schemata into one environment. Narrative engineers of all types and interests have and will create a plethora of interactive narratives as means to better understand the increasing complexity of the world and the social systems populating it. It is advisable to turn this into a joint effort.

---

[52] Evaluation examples related to interactor experiences in IDNs are: the game experience questionnaire [192], Roth's IDN experience evaluation framework [193], and quantitative based approaches towards validation [186, 194]. See also the work of the workinggroup 3 of the EU COST INDCOR project (https://indcor.eu/working-groups/wg3-evaluation/).
[53] In the context of experience the concept of retelling plays a significant role [198].
[54] Though it can be assumed that the collection of this data is critical, as similar approaches are already implemented in online gaming. What the collection of data regarding the authoring process means with respect to privacy issues concerning the narrative engineer (i.e. the perception of the performed work, established new work processes and hence ownership, etc) is unclear. A deep ethical discussion on various levels is certainly necessary (for narrative engineers as well as interactors).
[55] For particular articulation techniques, this might go as far as providing discussion environments between interactors and interactors with the narrative engineer. In particular journalistic environments might go into this direction (see Follow the Money (https://www.ftm.nl/) as an example). Technically this is not necessarily complicated, with respect to the engine-regulated interaction it is a challenge (it adds an additional level of complexity the narrative engineer has to cover).

Acknowledgements
This work would have not been possible with the kind support of the EU COST INNDCOR project.
I am in particular grateful to Hartmut Koenitz, Mirjam Palosaari-Elhadari, Sandy Louchart, Vincenco Lomardo, Mattia Bellini, Paul Groth, Bradley Allen, and Lise Stork for long and insightful discussions.

References

1. Dorst, Kees; Dijkhuis, Judith (1995). "Comparing paradigms for describing design activity". Design Studies. 16 (2): 261–274.
2. Asimow, M. (1962) Introduction to Design, Prentice-Hall, USA.
3. Archer, L. B. (1965) Systematic Method for Designers, The Design Council, UK
4. Norman, D. A. (1990): The Design of Everyday Things. New York, Doubleday
5. Simon, H.A. (1996). The sciences of the artificial. 3rd edition, MIT Press, Cambridge, MA, USA.
6. Preece, J., Rogers, Y. & Sharp, H. (2007). Interaction Design: Beyond Human-Computer Interaction, 2nd Edition, Chapter 9, John Wiley & Sons
7. Norman, D. A. (2009). THE WAY I SEE IT. Systems thinking: a product is more than the product. interactions 16, 5 (September + October 2009), 52–54.
8. Osborn, A. F. (1963) Applied Imagination: Principles and Procedures of Creative Thinking, Scribener's Sons, USA.
9. Schön, D.A. (1983). The reflective practitioner: How professionals think in action, Basic Books, USA.
10. Dorst, K.; Cross, Nigel (2001). Creativity in the design process: Co-evolution of problem–solution. Design Studies. 22 (5): 425–437.
11. Ralph, P. (2010). Comparing two software design process theories. International Conference on Design Science Research in Information Systems and Technology (DESRIST 2010), Springer, St. Gallen, Switzerland, pp. 139–153.
12. Cross, Nigel (2011). Design thinking : understanding how designers think and work. Berg.
13. Dorst, K. (2012). Frame Innovation: Create new thinking by design. Cambridge, MA: MIT Press
14. Koenitz, H., Di Pastena, A., Jansen, D., de Lint, B., Moss, A. (2018). The Myth of 'Universal' Narrative Models. In: Rouse, R., Koenitz, H., Haahr, M. (eds) Interactive Storytelling. ICIDS 2018. Lecture Notes in Computer Science(), vol 11318, pp. 107 – 120. Springer, Cham.
15. Aristotle (1968). Poetics, Introduction, Commentary and Appendixes by D.W. Lucas. Oxford: Oxford University Press.
16. Ricoeur, P. (1984 – 1988). Time and Narrative (Vol. 1 -3). The University of Chicago Press.
17. Bakhtin, M.M. (1981) The Dialogic Imagination: Four Essays by M.M. Bakhtin. Ed. Michael Holquist. Trans. Caryl Emerson and Michael Holquist. Austin and London: University of Texas Press
18. Propp, V. W. (1968). Morphology of the Folktale. University of Texas Press.
19. Lemon, L. T., & Reis, M. J. (1965). Russian Formalist Criticism - Four Essays. Lincoln: University of Nebraska Press.
20. Chomsky, N. (1965). Aspects of the theory of syntax. Cambridge, MA: MIT Press.
21. Lévi-Strauss, C. (1968). Structural Anthropology Volume 1, translated by Clair Jacobson, Brooke Grundfest Schoepf. London: Allen Lane.
22. Lévi-Strauss, C. (1977). Structural Anthropology Volume 2, translated by Monique Layton. London: Allen Lane.
23. Metz, C. (1974). Film Language: A Semiotic Of The Cinema. New York: Oxford University Press.
24. Price, G. (1973). A grammar of story: An introduction. The Hague: Mouton.
25. Greimas, J. (1983). Structural Semantics: An Attempt at a Method. Lincoln:University of Nebraska Press
26. Herman, D., Jahn, M., and Ryan, Marie-Laure (2007) Routledge Encyclopaedia of Narrative Theory, pp. 574 – 575. Taylor & Francis

27. van Dijk, T. (1972). Some aspects of text grammars: a study in theoretical linguistics and poetics. The Hague: Mouton.
28. Colby, B. N. (1973). A partial grammar of Eskimo folktales. American Anthropologist, 75, 645 - 62.
29. Lakoff, G. P. (1972). Structural complexity in fairy tales. The study of man, 1, 128 -190.
30. Mandler, J. M., &. Johnson, N. S. (1977). Remembrance of Things Parsed: Story Structure and Recall. Cognitive Psychology, 9, 111 - 151.
31. Rumelhart, D. E. (1977). Understanding and summarizing brief stories. In D. Laberge & S. J. Samuels (Eds.), Basic processes in reading: Perception and comprehension (pp. 265 - 303). Hillsdale, N.J.: Lawrence Erlbaum Associates.
32. Thorndyke, P. W. (1977). Cognitive structures in comprehension and memory of narrative discourse. Cognitive Psychology, 9, 77 - 100.
33. Knitsch, W., & van Dijk, T. A. (1978). Towards a model of text comprehension and production. Psychological review, 85, 363 -394.
34. Schank, R. C. (1972). Conceptual Dependency: A theory of natural language understanding. Cognitive Psychology, 3, 552 - 631.
35. Schank, R. C., & Abelson, R. (1977). Scripts, Plans, Goals And Understanding. Hillsdale, New Jersey: Lawrence Earlbaum Associates.
36. Schank, R. C. (1982). Dynamic memory. New York: Cambridge University Press.
37. Brewer, W. F. (1985). The story schema: Universal and culture-specific properties. In D. R. Olson, N. Torrance, & A. Hildyard (Eds.), Literacy, language, and learning (pp. 167–194). Cambridge, UK: Cambridge University Press.
38. Barthes, R. (1977). Image, Music, Text - Essays selected and translated by Stephen Heath. London: Fontana Press.
39. Iser, W. (19740. The Implied Reader: Patterns of Communication in Prose Fiction from Bunyan to Beckett. Baltimore: Johns Hopkins University Press.
40. Culler, J. (2001). The Pursuit of Signs: Semiotics, Literature, Deconstruction. London: Routledge.
41. Ronen, R. (1990). Paradigm Shift in Plot Models: An Outline of the History of Narratolog, Poetics Today, 11(4):817–842.
42. Sternberg, M. (1993). Expositional Modes and Temporal Ordering in Fiction. Bloomington: Indiana University Press
43. Chatman, S. (1978). Story and Discourse: Narrative Structure in Fiction and Film. New York: Ithaca
44. Campbell, J. (1949). The Hero with a Thousand Faces. Harper & Row, New York
45. McAdams, D.P. (1993). The Stories We Live By. New York: The Guilford Press
46. Bordwell, D. (1989). Making Meaning - Inference and Rhetoric in the Interpretation of Cinema. Cambridge, Massachusetts: Harward University Press.
47. Segre, C. (1979). Structure and Time - Narration, Poetry, Models. Chicago: The University of Chicago Press.
48. Anker, S. (2004), Real Writing with Readings: Paragraphs and Essays for College, Work, and Everyday Life (3rd ed.), Boston: Bedford Books
49. Crews, Frederick (1992), The Random House Handbook (6th ed.), New York: McGraw-Hill
50. Lodge, D. (1992). The Art of Fiction. Secker & Warburg
51. Freytag, G. (1894) Freytag's Technique of the Drama: An Exposition of Dramatic Composition and Art, translated and edited by Elias J. MacEwan. Chicago: Scott, Foresman and Company.
52. Jameson, F. (1998). Brecht and Method. London and New York: Verso,
53. Willett, John (1978). Art and Politics in the Weimar Period: The New Sobriety 1917–1933. New York: Da Capo Press
54. Bordwell, D. (1985). Narration in the Fiction Film. London: Methuen.
55. Bordwell, D. (1986). Classical Hollywood Cinema: Narrational Principles and Procedures. In P. Rosen (Eds.), Narrative, Apparatus, Ideology - A Film Theory Reader. New York: Columbia University Press
56. McCloud, S (2006). Making Comics: Storytelling Secrets of Comics, Manga, and Graphic Novels. William Morrow Paperbacks
57. Laban, R. (1950) The mastery of Movement on the stage. Macdonald and Evans.
58. Minden, E. G. (2007) The Ballet Companion: A Dancer's Guide, Simon and Schuster
59. Gloag, K and Beard, D. (2009) Musicology: The Key Concepts. New York: Routledge
60. Schell, J. (2008). The Art of Game Design. Amsterdam: Elsevier/Morgan Kaufmann.
61. Kintsch, W. (1988): The role of knowledge in discourse comprehension : a construction-integration model, Psychological Review, 1988, vol 95, pp. 163–182
62. Trabasso, T., van den Broek, P., & Suh, S. (1989). Logical necessity and transitivity of causal relations in stories. Discourse Processes,12,1–25.
63. Ozyurek, A., and Trabasso, T. (1997) ”Evaluation during the Understanding of Narratives.”. Discourse Processes, 23,:pp. 305-35
64. Zwaan, R. A., & Radvansky, G. A. (1998). Situation models in language comprehension and memory. Psychological Bulletin, 123(2), 162–185.
65. Scott-Rich, Shannon S., and Holly A. Taylor. "Not All Narrative Shifts Function Equally." Memory and Cognition, vol. 28, no. 7, 2000, pp. 1257–126
66. Nack, F. (1996) AUTEUR: The Application of Video Semantics and Theme Representation for Automated Film Editing Ph.D. Thesis, Lancaster University
67. Magliano, J. P., Loschky, L. C., Clinton, J. A., & Larson, A. M. (2013). Is reading the same as viewing? An exploration of the similarities and differences between processing text- and visually based narratives. In B.Miller, L. Cutting, & P. McCardle (Eds.),Unraveling the behavioral, neurobiological, and genetic components of reading comprehension(pp. 78–90). Baltimore, MD: Brookes.
68. Magliano, J. P., Clinton, J. A., O'Brien, E. J., & Rapp, D. N. (2018). Detecting differences between adapted narratives. In A. Dunst, J. Laubrock, & J. Wildfeuer (Eds.),Empirical comics research: Digital, multimodal, and cognitive methods(pp. 284–304). New York: Routledge.

69. Loschky, l.C., Adam M. Larson, Tim J. Smith, Joseph P. Magliano (2020). The Scene Perception & Event Comprehension Theory (SPECT) Applied to Visual Narrtives. topiCS 12(1): 311-351
70. Eco, U. (1977). A Theory of Semiotics. London: The Macmillan Press
71. Herman, D. (2013): Storytelling and the Sciences of Mind. Cambridge, Massachusetts: The MIT Press.
72. Zunshine, Lisa. 2006. Why We Read Fiction: Theory of Mind and the Novel. Theory and Interpretation of Narrative. Columbus: Ohio State University Press.
73. Bruni, L. S., Baceviciute, S., and Arief, M. (2014). 'Narrative Cognition in Interactive Systems: Suspense-Surprise and the P300 ERP Component'. In Proceedings of International Conference on Interactive Digital Storytelling (ICIDS), edited by Alex Mitchell, Clara Fernández-Vara, and David Thue, 164–75. Lecture Notes in Computer Science. Cham: Springer International Publishing.
74. Pope J. (2020). Further on down the digital road: Narrative design and reading pleasure in five New Media Writing Prize narratives. Convergence. 26(1):35-54.
75. Skains RL (2016). Creative Commons and Appropriation: Implicit Collaboration in Digital Works. Publications. 2016; 4(1):7.
76. The Arabian Nights (1992) based on the text of the fourteenth-century manuscript ed. by Muhsin Mahdi. Translated by Husain Haddawy. Everyman's Library.
77. Irwin, R. (2010). The Arabian Nights: A Companion. London: I.B. Tauris
78. Bruner, J. (1990). Acts of Meaning. Cambridge, MA: Harvard University Press.
79. Jakobson, R., & Halle, M. (1980). Fundamentals of Language. The Hague: Mouton Publishers.
80. Janowitz, M., & Street, D. (1966). The Social Organization of Education. In P. H. Rossi & B. J. Biddle (Eds.), The New Media and Education. Chicago: Aldine.
81. Manovich, L. (2003). "New Media From Borges to HTML". The New Media Reader. Ed. Noah Wardrip-Fruin & Nick Montfort. Cambridge, Massachusetts.
82. Manovich, L. (2001). The Language of New Media. Cambridge : MIT Press.
83. Wiener, N.(1950). Cybernetics and Society: The Human Use of Human Beings. Houghton Mifflin
84. Foerster, H. von (1979). "Cybernetics of cybernetics". In K. Krippendorff, ed., Communication and Control in Society, New York: Gordon and Breach, pp 5– 8.
85. Foerster, H.von (2003). Understanding Understanding: Essays on Cybernetics and Cognition. New York: Springer.
86. Glasersfeld, Ernst von. (1984). An introduction to radical constructivism. In P. Watzlawick (Ed.), The invented reality. Norton.
87. Murray, J. H. (1997). Hamlet on the Holodeck. The MIT Press
88. Murray, J. (May 1, 2004). "From Game-Story to Cyberdrama". Electronic Book Review. https://electronicbookreview.com/essay/from-game-story-to-cyberdrama/ Retrieved: 15/04/2022
89. Scheff, T. J. (1979). Catharsis in Healing, Ritual, and Drama. University of California Press
90. Aarseth, Espen J. (1997). Cybertext—Perspectives on Ergodic Literature. Johns Hopkins University Press.
91. Nelson, T. H (2003). "A File structure for the Complex, the Changing, and the Indeterminate". The New Media Reader. Ed. Noah Wardrip-Fruin & Nick Montfort. Cambridge, Massachusetts, pp. 133 -148.
92. Moulthrop, S. (2003). "You Say You Want a Revolution? Hypertext and the law of media.”". The New Media Reader. Ed. Noah Wardrip-Fruin & Nick Montfort. Cambridge, Massachusetts. Pp. 692 - 704
93. Coover, R., S. (2003). "The End of Books”. The New Media Reader. Ed. Noah Wardrip-Fruin & Nick Montfort. Cambridge, Massachusetts. Pp. 705 – 709.
94. Ryan, M-L; Emerson, L.; and Robertson, B. J. (2014). The Johns Hopkins Guide to Digital Media. Baltimore, MD: JHU Press. p. 262.
95. Peters, J. D. (1999). Speaking into the air : a history of the idea of communication. Chicago: University of Chicago Press.
96. Littlejohn, S.; Foss, K. (2009), "Definitions of Communication", Encyclopedia of Communication Theory, Thousand Oaks: SAGE Publications, Inc., pp. 296–299,
97. Juul, J. (2001). A clash between game and narrative". https://www.jesperjuul.net/thesis/ Retrieved 02.05.2022.
98. Juul, J. (2001). "Games Telling Stories: A brief note on games & Narratives". Games Studies. 1 (1). http://www.gamestudies.org/0101/juul-gts/ retrieved: 02/05/2022
99. Juul, J. (2006). Half-Real: Video Games between Real Rules and Fictional Worlds. MIT Press.
100. Aarseth, Espen (2001). "Computer Game Studies, Year One". Game Studies. 1 (1). http://gamestudies.org/0101/editorial.html retrieved 02.05.2022
101. Wardrip-Fruin, N. and Harrigan, P. (2004). First Person - New Media as Story, Performance, and Game. MIT Press
102. Harrigan, P. and Wardrip-Fruin, N (2007). Second Person - Role-Playing and Story in Games and Playable Media, MIT Press
103. Harrigan, P. and Wardrip-Fruin, N (2009).Third Person - Authoring and Exploring Vast Narratives. MIT Press
104. Laurel, B. (1991). Computers as Theatre, Addison-Wesley
105. Bates, J. (1992). The Nature of Characters in Interactive Worlds and the Oz Project (Technical Report No. CMU-CS-92-200). School of Computer Science Carnegie Mellon University.
106. Kelso, M, Weyhrauvch, and Bates, J. (1993). Dramatic presence. In PRESENCE, The Journal of Teleoperators and Virtual Environments.MIT Press, 2(1), pp 1-15
107. Bates, J. (1994). The role of Emotion in Believable Agents. Communications of the ACM, 37(7), 122 - 125.
108. Perlin K. and Goldberg, A. (1996). Improv: a system for scripting interactive actors in virtual worlds. In Proceedings of the 23rd annual conference on Computer graphics and interactive techniques (SIGGRAPH '96). Association for Computing Machinery, New York, NY, USA, 205–216.

109. Hayes-Roth, B. (1995). Agents on Stage: Advancing the State of the Art of AI. In C. S. Melish (Ed.), IJCAI-95 International Joint Conference on Artificial Intelligence, (pp. 967 - 971). Montréal, Canada: August 10 - 25, 1995.
110. Hayes-Roth, B., Sincoff, E., Brownston, L., Huard, R., & Lent, B. (1994). Directed Improvisation (Technical Report No. KSL-94-61). Stanford University.
111. Peinado, F., Cavazza, M., Pizzi., D. 2008. Revisiting Character-Based Affective Storytelling under a Narrative BDI Framework. In the Proceedings of First Joint Conference on Interactive Digital Storytelling (ICIDS), Erfurt, Germany, November 2008: pp.83-88
112. Aylett, R.,Louchart, S. and Weallans, A. : 'Research in Interactive Drama Environments, Role-Play and Story-Telling', in Si, M., Thue, D.,André, E., Lester, J. L., Tanenbaum, J. (eds). 4th International Conference, ICIDS 2011, pp. 1-12, LNCS, vol 7069. Springer, Heidelberg (2011)
113. Bostan B, Marsh T (2012) Fundamentals of interactive storytelling. Acad J Inform Technol 3(8):20–42. Last accessed: 10.05.2022 https://www.researchgate.net/publication/268978507_Fundamentals_Of_Interactive_Storytelling
114. Mateas, M. and Stern, A. , "Structuring Content in the Façade Interactive Drama Architecture." Artificial Intelligence and Interactive Digital Entertainment (AIIDE 2005), Marina del Rey, CA. June 1-3, 2005.
115. Antin, D. (2020-05-05). "The Technology of the Metaverse, It's Not Just VR". The Startup. https://medium.com/swlh/the-technology-of-the-metaverse-its-not-just-vr-78fb3c603fe9 last accessed: 03.05.2022
116. Koenitz. H (2023). Understanding Interactive Digital Narrative - Immersive expressions for a complex time. Routledge.
117. Koenitz, H. (2015). Towards a Specific Theory of Interactive Digital Narrative. In *Interactive Digital Narrative* (pp. 107-121). Routledge.
118. Roth, C., Van Nuenen, T., & Koenitz, H. (2018, December). Ludonarrative Hermeneutics: A Way Out and the Narrative Paradox. In International Conference on Interactive Digital Storytelling (pp. 93-106). Springer, Cham.
119. Spierling, U and Szilas, N.: 'Authoring Issues beyond Tools', in Iurgel, I. A., Zagalo, N., Petta, P. (eds). 2nd International Conference, ICIDS 2009, pp. 50 - 61, LNCS, vol 5915. Springer, Heidelberg (2009).
120. Hardman, L., Obrenovic, Z., Nack, F., Kerherve, B., and Piersol, K. (2008) Canonical Processes of Semantically Annotated Media Production. Special Issue on 'Canonical Process of Media Production', Multimedia Systems Journal, 14(6), pp. 327 – 340.
121. Koenitz, H. (2015). Design Approaches for Interactive Digital Narrative. In: Schoenau-Fog, H., Bruni, L., Louchart, S., Baceviciute, S. (eds) Interactive Storytelling. ICIDS 2015. Lecture Notes in Computer Science(), vol 9445. Springer, Cham. Pp. 50 - 57
122. Swartjes, I., Theune, M. (2009). Iterative Authoring Using Story Generation Feedback: Debugging or Co-creation?. In: Iurgel, I.A., Zagalo, N., Petta, P. (eds) Interactive Storytelling. ICIDS 2009. Lecture Notes in Computer Science, vol 5915, pp. 62 – 73. Springer, Berlin, Heidelberg
123. Tudor, A. (1974). Image And Influence. London: George Allen & Unwin Ltd.
124. Dance, F. E. (1967). A helical model of communication. Human Communication Theory, New York, NY: Holt, Rinehart and Winston..
125. Hayles, N. Katherine. (2005). My mother was a computer: digital subjects and literary texts. Chicago: University of Chicago Press.
126. Knoller, N., Roth, C., Haak, D. (2021). The Complexity Analysis Matrix. In: Mitchell, A., Vosmeer, M. (eds) Interactive Storytelling. ICIDS 2021. Lecture Notes in Computer Science (), vol 13138. Pp. 478 -487. Springer, Cham.
127. C. David Mortensen (1972). Communication: The Study of Human Communication, New York: McGraw-Hill Book Co. see Chapter 2, "Communication Models."
128. McNamara, J. (1994). From Dance to Text and Back to Dance: A Hermeneutics of Dance Interpretive Discourse, PhD thesis, Texas Woman's University.
129. Grondin, J. (1994). Introduction to Philosophical Hermeneutics. Yale University Press.
130. Heidegger, M. (1971). "The Origin of the Work of Art." Poetry, Language, Thought. Trans. Albert Hofstadter. NY: Harper Collins.
131. Gadamer, H. (2018). Hermeneutical Foundations (see 2.2 Ontology and Hermeneutics). Stanford Encyclopedia of Philosophy. http://seop.illc.uva.nl/entries/gadamer/ last accessed: 01.05.2022
132. B. Kybartas and R. Bidarra (2017), "A Survey on Story Generation Techniques for Authoring Computational Narratives," in IEEE Transactions on Computational Intelligence and AI in Games, vol. 9, no. 3, pp. 239-253.
133. Shibolet, Y., Knoller, N., Koenitz, H. (2018). A Framework for Classifying and Describing Authoring Tools for Interactive Digital Narrative. In: Rouse, R., Koenitz, H., Haahr, M. (eds) Interactive Storytelling. ICIDS 2018. Lecture Notes in Computer Science(), vol 11318. Springer, Cham.pp. 523 -533
134. Kitromili, S., Jordan, J. and Millard, D. E. (2019). What is Hypertext Authoring? In Proceedings of the 30th ACM Conference on Hypertext and Social Media (HT '19). Association for Computing Machinery, New York, NY, USA, 55–59
135. EU Cost action INDCOR (2022): List of IDN authoring tools: https://omeka-s.indcor.eu/s/idn-authoring-tools/item-set/43
136. Polle T. Zellweger, Anne Mangen, and Paula Newman. 2002. Reading and writing fluid Hypertext Narratives. In Proceedings of the thirteenth ACM conference on Hypertext and hypermedia (HYPERTEXT '02). Association for Computing Machinery, New York, NY, USA, 45–54.
137. Claus Atzenbeck, Mark Bernstein, Marwa Ali Al-Shafey, and Stacey Mason. 2013. TouchStory: combining hyperfiction and multitouch. In Proceedings of the 24th ACM Conference on Hypertext and Social Media (HT '13). Association for Computing Machinery, New York, NY, USA, 189–195.
138. Ventura, D., Brogan, D. (2003). Digital Storytelling with DINAH: Dynamic, Interactive, Narrative Authoring Heuristic. In: Nakatsu, R., Hoshino, J. (eds) Entertainment Computing. IFIP — The International Federation for Information Processing, vol 112. Springer, Boston, MA

139. Caropreso, Maria Fernanda; Inkpen, Diana; Keshtkar, Fazel; Khan, Shahzad (2012).Template Authoring Environment for the Automatic Generation of Narrative Content. Journal of Interactive Learning Research, v23 n3 p227-249.
140. James Ryan, Ethan Seither, Michael Mateas, and Noah Wardrip-Fruin. 2016. Expressionist: An Authoring Tool for In-Game Text Generation. In Interactive Storytelling: 9th International Conference on Interactive Digital Storytelling, ICIDS 2016, Los Angeles, CA, USA, November 15–18, 2016, Proceedings. Springer-Verlag, Berlin, Heidelberg, 221–233.
141. Louchart, S., Swartjes, I., Kriegel, M., Aylett, R. (2008). Purposeful Authoring for Emergent Narrative. In: Spierling, U., Szilas, N. (eds) Interactive Storytelling. ICIDS 2008. Lecture Notes in Computer Science, vol 5334. Springer, Berlin, Heidelberg.
142. Pizzi, D., Cavazza, M. (2008). From Debugging to Authoring: Adapting Productivity Tools to Narrative Content Description. In: Spierling, U., Szilas, N. (eds) Interactive Storytelling. ICIDS 2008. Lecture Notes in Computer Science, vol 5334. Springer, Berlin, Heidelberg. https:
143. Aylett, R., Louchart, S., Weallans, A. (2011). Research in Interactive Drama Environments, Role-Play and Story-Telling. In: Si, M., Thue, D., André, E., Lester, J.C., Tanenbaum, T.J., Zammitto, V. (eds) Interactive Storytelling. ICIDS 2011. Lecture Notes in Computer Science, vol 7069. Springer, Berlin, Heidelberg.
144. Szilas, N., Richle, U. and Dumas, J. E. : 'Structural Writing, a Design Principle for Interactive Drama', in Oyarzun, D., Peinado, F., Young, R. M., Elizalde, A., Méndez, G. (eds). 5th International Conference, ICIDS 2012, pp. 72–83, LNCS, vol 7648. Springer, Heidelberg (2012).
145. Lombardo, V., Damiano, R. Semantic annotation of narrative media objects. Multimed Tools Appl 59, 407–439 (2012).
146. Ryan, J.O., Mateas, M., Wardrip-Fruin, N. (2015). Open Design Challenges for Interactive Emergent Narrative. In: Schoenau-Fog, H., Bruni, L., Louchart, S., Baceviciute, S. (eds) Interactive Storytelling. ICIDS 2015. Lecture Notes in Computer Science(), vol 9445. Springer, Cham.
147. Stacey Mason, Ceri Stagg, and Noah Wardrip-Fruin. 2019. Lume: a system for procedural story generation. In Proceedings of the 14th International Conference on the Foundations of Digital Games (FDG '19). Association for Computing Machinery, New York, NY, USA, Article 15, 1–9.
148. Jacob Garbe, Max Kreminski, Ben Samuel, Noah Wardrip-Fruin, and Michael Mateas. 2019. StoryAssembler: an engine for generating dynamic choice-driven narratives. In Proceedings of the 14th International Conference on the Foundations of Digital Games (FDG '19). Association for Computing Machinery, New York, NY, USA, Article 24, 1–10.
149. Brooks KM (1999). Metalinear Cinematic Narrative: Theory, Process, and Tool. MIT Ph.D. Thesis
150. Davenport G; Bradley B; Agamanolis S; Barry B; Brooks KM (2000). Synergistic storyscapes and constructionist cinematic sharing. IBM Systems Journal, vol.39, no.3-4, 2000, pp.456-69.
151. Falkovych, K., Nack, F. Context aware guidance for multimedia authoring: harmonizing domain and discourse knowledge. Multimedia Systems 11, 226–235 (2006)
152. Bocconi, S. (2006). Vox Populi: generating video documentaries from semantically annotated media repositories. PhD thesis from the Technical University of Eindhoven.
153. Koenitz, H., Chen, KJ. (2012). Genres, Structures and Strategies in Interactive Digital Narratives – Analyzing a Body of Works Created in ASAPS. In: Oyarzun, D., Peinado, F., Young, R.M., Elizalde, A., Méndez, G. (eds) Interactive Storytelling. ICIDS 2012. Lecture Notes in Computer Science, vol 7648. Springer, Berlin, Heidelberg.
154. Marian F. Ursu, Maureen Thomas, Ian Kegel, Doug Williams, Mika Tuomola, Inger Lindstedt, Terence Wright, Andra Leurdijk, Vilmos Zsombori, Julia Sussner, Ulf Myrestam, and Nina Hall. 2008. Interactive TV narratives: Opportunities, progress, and challenges. ACM Trans. Multimedia Comput. Commun. Appl. 4, 4, Article 25 (October 2008), 39 pages.
155. Marian F. Ursu, Vilmos Zsombori, John Wyver, Lucie Conrad, Ian Kegel, and Doug Williams. 2009. Interactive documentaries: A Golden Age. Comput. Entertain. 7, 3, Article 41 (September 2009), 29 pages.
156. Dick C. A. Bulterman, Pablo Cesar, and Rodrigo Laiola Guimarães. 2013. Socially-aware multimedia authoring: Past, present, and future. ACM Trans. Multimedia Comput. Commun. Appl. 9, 1s, Article 35 (October 2013), 23 pages.
157. Guimarães, R. L. 2014. Socially-Aware Multimedia Authoring. Doctoral thesis. VU Amsterdam.
158. Iván Otero-Gonzále and Jorge Vázquez-Herrero (2022). Open and commercial tools to generate a digital interactive story: systematic review and features analysis. New Review of Hypermedia and Multimedia, to be published.
159. Samuel, B., Mateas, M., Wardrip-Fruin, N.: The design of writing buddy: a mixed-initiative approach towards computational story collaboration. In: Nack, F., Gordon, A.S. (eds.) ICIDS 2016. LNCS, vol. 10045, pp. 388–396. Springer, Cham (2016). https://doi.org/10.1007/978-3-319-48279-8 34
160. Jemily Rime, Alan Archer-Boyd, and Tom Collins. 2023. How Will You Pod? Implications of Creators' Perspectives for Designing Innovative Podcasting Tools. ACM Trans. Multimedia Comput. Commun. Appl. 20, 3, Article 69 (March 2024), 25 pages. https://doi.org/10.1145/3625099
161. Mina Lee et al. 2024. A Design Space for Intelligent and Interactive Writing Assistants. In Proceedings of the 2024 CHI Conference on Human Factors in Computing Systems (CHI '24). Article 1054, 1–35. https://doi.org/10.1145/3613904.3642697
162. Serbanescu, Anca. (2024). Human-AI system Co-creativity. Understanding the relationship between Designer and AI systems in the field of Interactive Digital Narrative. PhD thesis, School of design, Politecnico di Milano, Italy.
163. Luhman, N. (1996). Social Systems. Translated by John Bednarz, Jr. with Dirk Baecker. Standford University Press.
164. Koenitz. H.; Palosaari-Elhadari, M.; Louchart, S. & Nack, F. (2020). INDCOR white paper 1: A shared vocabulary for IDN (Interactive Digital Narratives). October 20, White Paper, COST Action 18230 (https://arxiv.org/abs/2010.10135)
165. Peng, C., Xia, F., Naseriparsa, M. et al. Knowledge Graphs: Opportunities and Challenges. Artificial Intelligence Review 56, 13071–13102 (2023). https://doi.org/10.1007/s10462-023-10465-9

166. Yu, H., & Riedl, M.O. (2013). Toward personalized guidance in interactive narratives. International Conference on Foundations of Digital Games. https://www.semanticscholar.org/paper/Toward-personalized-guidance-in-interactive-Yu-Riedl/34eff4e12197f53f7909986df03c63a2ba1a6afc, last accessed 2024/03/22
167. Singh, K., Lytra, I., Sethupat, A., Shekarpour, S., Vidal, M.E., Lehmann, J.: No one is perfect: Analysing the performance of question answering components over the DBpedia knowledge graph. Journal of Web Semantics 65 (2020). https://doi.org/10.1016/j.websem.2020.100594
168. Hogan, A., Blomqvist, E., Cochez, M., d'Amato, C., de Melo, G., Gutierrez, C., Labra Gayo, J.E., Kirrane, S., Neumaier, S., Polleres, A., Navigli, R., Ngomo, A.-C., Rashid, S.M., Rula, A., Schmelzeisen, L., Sequeda, J., Staab, S.,Zimmermann, A.: Knowledge Graphs. ACM Computing Surveys 54(4), 1–37 (2022). https://doi.org/10.1145/3447772
169. Abhilash, M., Nack, F. (2025). Design of Knowledge Graphs for Interactive Digital Narratives Authoring. In: Murray, J.T., Reyes, M.C. (eds) Interactive Storytelling. ICIDS 2024. Lecture Notes in Computer Science, vol 15468. Springer, Cham. https://doi.org/10.1007/978-3-031-78450-7_18
170. Paranich, J. (2024). The Role of Intent in the UI of IDN Authoring. MSc thesis. University of Amsterdam.
171. Cevey, s., Matt Chadburn, Tom Leitch, Michael Smethurst, Helen Lippell, Dan Brickley, and Paul Rissen. 2013. Storyline Ontology, An ontology to represent News Storylines. Retrieved March 17, 2021 from https://www.bbc.co.uk/ ontologies/storyline
172. Varadarajan, U. and Dutta, B. (2020). Models for Narrative Information: A Study. (Available from https:/ https://arxiv.org/abs/2110.02084/arxiv.org/) last accessed:
173. Zarah Saied (2021). Knowledge representation in IDN authoring tools. MSC Thesis. Master Information Studies, Informatics Institute, Faculty of Science, University of Amsterdam
174. Troncy, R., Bartosz Malocha and André Fialho.(2010) Linking. Events with Media. In the Open Track of the Linked Data Triplification Challenge, colocated with the 6th International Conference on Semantic Systems (I-SEMANTICS'10), Graz, Austria, September 1-3 10.1145/1839707.1839759.
175. Damiano, R., Lombardo, V., & Pizzo, A. (2019). The ontology of drama. Appl. Ontology, 14,(1), 79-118.
176. Meghini, C., Valentina Bartalesi, and Daniele Metilli. 2021. Representing narratives in digital libraries: The narrative ontology. Semantic Web Preprint (2021), 1–24.
177. Stephen Toulmin (1958). The Uses of Argument (1958) 2nd edition 2003, Cambridge University Press
178. Hinton, M., Wagemans, J.H.M. (2022). Evaluating Reasoning in Natural Arguments: A Procedural Approach. Argumentation 36, 61–84. https://doi.org/10.1007/s10503-021-09555-1
179. Jean H.M. Wagemans (2023). How to identify an argument type? On the hermeneutics of persuasive discourse. Journal of Pragmatics, Volume 203, Pages 117-129, https://doi.org/10.1016/j.pragma.2022.11.015.
180. Ahmady, A.: Supporting Authorial Creativity in Interactive Storytelling: GPT-4's Role in Text-Based Interactive Fiction. University of Amsterdam (2024)
181. Koenitz, H., Eladhari, M.P., Barbara, J. (2025). Can AI Create an Interactive Digital Narrative? A Benchmarking Framework to Evaluate Generative AI Tools for the Design of IDNs. In: Murray, J.T., Reyes, M.C. (eds) Interactive Storytelling. ICIDS 2024. Lecture Notes in Computer Science, vol 15467. Springer, Cham. https://doi.org/10.1007/978-3-031-78453-8_11
182. Knight, Y., Eladhari, M.P. (2025). Mixed Initiative Comic Making in an Artistic Practice. In: Murray, J.T., Reyes, M.C. (eds) Interactive Storytelling. ICIDS 2024. Lecture Notes in Computer Science, vol 15467. Springer, Cham. https://doi.org/10.1007/978-3-031-78453-8_10
183. Johnson-Bey, S. et al. (2025). Building Visual Novels with Social Simulation and Storylets. In: Murray, J.T., Reyes, M.C. (eds) Interactive Storytelling. ICIDS 2024. Lecture Notes in Computer Science, vol 15468. Springer, Cham. https://doi.org/10.1007/978-3-031-78450-7_9
184. Nack, F. & Putz, W. (2001) Designing Annotation Before It's Needed. In Proceedings of the 9th ACM International Conference on Multimedia, pp. 251 – 260, Ottawa, Canada, Sept. 30 – Oct. 5, 2001. to appear in ACM MM 2001 – Ottawa, Canada, Sept. 30-Oct. 5, 2001.
185. Bulterman, D. (2018). SMIL: Synchronized multimedia integration language. In MediaSync: Handbook on Multimedia Synchronization. (pp. 359–385). doi:10.1007/978-3-319-65840-7_13
186. Papilaya, D. and Nack, F (2022) An investigation on the usability of socio-cultural features for the authoring support during the development of Interactive Discourse Environments (IDE). In: Vossmer, M., Holloway-Attaway, L. (eds), Interactive Storytelling – 15th International Conference on Interactive Storytelling, ICIDS 2022, LNCS, vol. 13762, pp. 309 - 328. Springer, Lecture Notes in Computer Science
187. Akkerman, S., Nack, F. (2023). Figure of Speech Detection and Generation as a Service in IDN Authoring Support. In: Holloway-Attaway, L., Murray, J.T. (eds) Interactive Storytelling. ICIDS 2023. Lecture Notes in Computer Science, vol 14384. Springer, Cham. https://doi.org/10.1007/978-3-031-47658-7_8
188. Latour, Bruno (1987). Science in action: how to follow scientists and engineers through society. Cambridge, Massachusetts: Harvard University Press. ISBN 978-0-674-79291-3.
189. Pasquetto, I. V., Borgman, C. L., & Wofford, M. F. (2019). Uses and Reuses of Scientific Data: The Data Creators' Advantage. Harvard Data Science Review, 1(2). https://doi.org/10.1162/99608f92.fc14bf2d
190. Daniel Andrews and Chris Baber. 2014. Visualizing interactive narratives: employing a branching comic to tell a story and show its readings. In Proceedings of the SIGCHI Conference on Human Factors in Computing Systems (CHI '14). Association for Computing Machinery, New York, NY, USA, 1895–1904.
191. Paul De Bra, Natalia Stash, Wouter Boereboom, Celine Chen, Joris Den Ouden, Martijn Kunstman, John Oostrum, and Egon Verbakel. 2016. ALAT: Finally an Easy To Use Adaptation Authoring Tool. In Proceedings of the 27th ACM Conference on Hypertext and Social Media (HT '16). Association for Computing Machinery, New York, NY, USA, 213–218. https://doi.org/10.1145/2914586.2914627

192. IJsselsteijn, W., Poels, K., de Kort, Y.: The Game Experience Questionnaire: Development of a self-report measure to assess player experiences of digital games. TU Eindhoven (2008). https://research.tue.nl/en/publications/the-game-experience-questionnaire
193. Roth, C.(2016) Experiencing interactive storytelling. PHD. Thesis. Vrije Universiteit Amsterdam https://research.vu.nl/en/publications/experiencing-interactive-storytelling.
194. Roth, C. and Koenitz, H. (2017). Towards Creating a Body of Evidence-based Interactive Digital Narrative Design Knowledge: Approaches and Challenges. In Proceedings of the 2nd International Workshop on Multimedia Alternate Realities (AltMM '17). Association for Computing Machinery, New York, NY, USA, 19–24. https://doi.org/10.1145/3132361.3133942
195. Phillips, P. J; Hahn, C. A.; Fontana, P. C.; Yates, . N.; Greene, K.; Broniatowski, D. A.; Przybocki, M. A. (2021). "Four Principles of Explainable Artificial Intelligence". doi:10.6028/nist.ir.8312. https://nvlpubs.nist.gov/nistpubs/ir/2021/NIST.IR.8312.pdf last accessed: 23.5.2022
196. Vilone, Giulia, and Luca Longo. 2021. "Classification of Explainable Artificial Intelligence Methods through Their Output Formats" Machine Learning and Knowledge Extraction 3, no. 3: 615-661. https://doi.org/10.3390/make3030032
197. Bellini, M. (2025). The Complexity of Video Games. PhD thesis. University of Tartu, Institute of Cultural Research Research group on Narrative, Culture, and Cognition
198. Eladhari, M.P (2018) Re-Tellings: the fourth layer of narrative as an instrument for critique. In: Rouse, R., Koenitz, H., Haahr, M. (eds.) ICIDS 2018. LNCS, vol. 11318, pp. 65–78. Springer, Cham (2018). https://doi.org/10.1007/978-3-030-04028-4 5